\documentclass[reprint,superscriptaddress,showpacs,nofootinbib,amsmath,amssymb,aps,prb]{revtex4-1}
\usepackage{graphicx}
\usepackage{dcolumn}
\usepackage{bm}
\usepackage[utf8]{inputenc}
\usepackage[frenchb]{babel}
\usepackage{braket}
\usepackage{xcolor}
\usepackage[T1]{fontenc}

\begin{document}

\preprint{APS/123-QED}

\title{Controllable fusion of electromagnetic bosons in two-dimensional semiconductors}

\author{S. V. Andreev}
\email[Electronic adress: ]{Serguey.Andreev@gmail.com}
\affiliation{Physikalisches Institut, Albert-Ludwigs-Universit\"at Freiburg, Hermann-Herder-Strasse 3, 79104 Freiburg, Germany}

\date{\today}

\begin{abstract}

We propose a physical principle for implementation of controllable interactions of identical electromagnetic bosons (excitons or polaritons) in two-dimensional (2D) semiconductors. The key ingredients are tightly bound biexcitons and in-plane anisotropy of the host structure due to, \textit{e.g.}, a uniaxial strain. We show that  anisotropy-induced splitting of the radiative exciton doublet couples the biexciton state to continua of boson scattering states. As a result, two-body elastic scattering of bosons may be resonantly amplified when energetically tuned close to the biexciton by applying a transverse magnetic field or tuning the coupling with the microcavity photon mode. At the resonance bosonic fields undergo quantum reaction of fusion accompanied by their squeezing. For excitons, we predict giant molecules (Feshbach dimers) which can be obtained from a biexciton via rapid adiabatic sweeping of the magnetic field across the resonance. The molecules possess non-trivial entanglement properties. Our proposal holds promise for the strongly-correlated photonics and quantum chemistry of light. 

\end{abstract}


\maketitle

\section{Introduction}

\subsection{Foreword}

Semiconductor optics holds great promise for the practical implementation of the emergent quantum technologies \cite{Lodahl2015}. The central issues are entanglement and squeezing \cite{Scully}, and these have put bound exciton pairs into the focus of applied research over the past decades \cite{Miller, Sie2015, XXinTMD1, Hao2017, XXinTMD3, Thouin2018, Andreev2016, Andreev2020}. The state-of-the-art embodies entangled photon pairs from biexcitons in nanocrystals \cite{Tamarat2020} and quantum dots (QD's) \cite{Stevenson2006, Salter2010, Dousse2010}, and the low-threshold biexciton lasing in two-dimensional (2D) heterostructures \cite{Kondo1998, Grim2014, Booker2018}.

In contrast to "zero-dimensional" states in QD's, 2D excitons represent propagating (guided) electromagnetic waves,\cite{de2001evanescent} which possess massive dispersion and may undergo two-body quantum scattering due to Coulomb forces between the constituent electrons and holes.\cite{Ivanov1998} The exciton dispersion may be efficiently tailored by photonic engineering of the environment,\cite{Fang2019} with the ultimate case being the microcavity exciton-polariton with drastically reduced effective mass.\cite{Weishbuch1992} At low densities 2D excitons behave as interacting bosons and may accumulate in a single quantum state featuring the phenomena of Bose-Einstein condensation (BEC) and superfluidity. Quantum collective effects associated with BEC and pairing of 2D excitons have been the subject of the experimental \cite{Wolfe, High2012, Gorbunov2012, BarJoseph, Dubin, Knorr2020, Sigl2020} and theoretical studies \cite{Hanamura, Noziers, Ivanov1998, Schuck1998, Lozovik2002, Andreev2016, Hanai2017}. Natural connection of these low-temperature equilibrium phenomena to lasing has been a central problem in polaritonics.\cite{Chiocchetta2017}        

For the quantum optical applications, 2D heterostructures have several noteworthy benefits as compared to QD's. Delocalized 2D ensembles are less prone to the Auger recombination known to be deleterious for lasers \cite{Grim2014}. Many-body states of interacting excitons featuring macroscopic entanglement and squeezing may also outperform the isolated localized pairs in such applications as quantum optical lithography \cite{Boto2000}, sensing \cite{Dowling2008} and all-optical quantum computing \cite{Takahiro2021}.

Aiming primarily at discussion of the underlying two-body physics, we shall use a collective term "electromagnetic bosons" referring either to 2D excitons or polaritons from a single perspective. An electromagnetic boson is essentially a photon dressed by electronic excitations in a 2D semiconductor medium with direct band gap. We shall not be concerned with the so-called "gray" and "dark" excitons.\cite{Robert2017} As such, electromagnetic bosons are statistically identical quasiparticles.     

A steppingstone for fundamental research on mixtures of electromagnetic bosons and their molecules, as well as for their practical use, would be \textit{in situ} control over the inter-conversion process
\begin{equation}
\label{ResonantReaction}
\alpha_\sigma+\alpha_{\sigma^\prime}\leftrightarrows \alpha_{\uparrow}^\prime\alpha_\downarrow^\prime,
\end{equation}
where $\alpha_\sigma$ is either an exciton ($\alpha=X$) or a polariton ($\alpha=L$) with spin $\sigma$ ("$\uparrow$" or "$\downarrow$") and $\alpha_{\uparrow}\alpha_\downarrow$ is a molecule (necessarily spin-singlet, as discussed in Section \ref{Section2}). For atomic Bose (as well as Fermi) gases, such control has taken the name of Feshbach resonance.\cite{Chin2010} Controllable synthesis of atomic molecules has spurred development of ultra-cold quantum chemistry \cite{Bell2009} and "super-chemistry" exploiting bosonic stimulation of reactions in the presence of BEC.\cite{Heinzen2000, Moore2002, Zhang2023}

The Feshbach resonance concurrently offers a mechanism to adjust the strength of two-body interactions between particles at will, from increasingly strong repulsion to attraction through unitarity. \cite{Chin2010} This is achieved by tuning the energy of the scattering channel of interest (called "open" channel) with respect to the energy of the molecule. The scattering channel admitting the molecular state is designed as energetically "closed" by sustaining the corresponding dissociation threshold above the energy of colliding particles in the "open" channel. The Feshbach resonance occurs when the "open" channel crosses the molecular level. Provided the molecule is tightly bound, the energy mismatch between the channels is large, making the \textit{coherent coupling} to the continuum of the "closed" channel ineffective. On the other hand, coupling to the resonant molecular state affects dramatically particle scattering in the "open" channel. The nature of the coherent coupling depends on the particular setting and will be discussed in more detail below.

Such an additional possibility of tunable interactions would be of paramount interest in the context of electromagnetic bosons, as it would enable strong photonic non-linearities at ultra-low densities. On general grounds, one would also expect strong interactions and pairing to be accompanied by entanglement and squeezing of light.\cite{Scully} The latter features have not yet been fully appreciated in the context of ultra-cold atoms. In contrast, entanglement and squeezing emerge naturally in non-linear optics of 2D semiconductors, where interactions between electromagnetic bosons provide non-linear susceptibilities.\cite{Boyd2008, Takayama2002, Schumacher2007}

\subsection{Coherent link}         
 
The crucial ingredient of a Feshbach resonance is a coherent link between the "open" scattering channel of interest and the "closed" molecular channel. In prototypical atomic Feshbach resonances such a link has been enabled either by the hyperfine interaction \cite{Timmermans1999} or by the photon exchange with an external radiation field \cite{Fedichev}. In semiconductors, signatures of a Feshbach resonance have recently been reported for a twisted bilayer 2D heterostructure \cite{Schwartz2021} and attributed to hybridization of exciton-electron scattering states with the intralayer (closed channel) trion state \cite{Kuhlenkamp2022}. In this case the coherent link is presumably due to interlayer electron (or hole) tunnelling.   

For a pair of identical electromagnetic bosons, two possible mechanisms of coherent coupling to a biexciton have been proposed. The first one relies on the so-called giant oscillator strength model \cite{Rashba1962, Ivanov1998} of a biexciton put into a microcavity: the biexciton may coherently dissociate into an exciton and a microcavity photon \cite{carusotto2010}. Several theoretical \cite{Wouters2007, Bastarrachea-Magnani2019, Denning2022} and experimental \cite{Navadeh-Toupchi2019, Takemura2014, Takemura2017} studies have been carried out along this direction. The second proposal exploits the dipolar repulsion between the excitons electrically polarized in the transverse direction: the biexciton is thus transformed into a shape resonance. \cite{Andreev2016, Andreev2020} The latter mechanism applies equally well for microcavity polaritons and bare excitons (including also their "gray" and "dark" species). Two recent independent experiments \cite{Rosenberg2018, Datta2022} have reported divergence of the dipolar polariton blueshift consistent with a broad shape resonance. \cite{Andreev2020} Remarkably, the theory \cite{Andreev2020} predicts efficient squeezing of the light emitted by polaritons interacting via such a broad resonance. While viability of these proposals may be subjected to debates and further experimental tests, technological relevance of the direction calls for the quest of their possible alternatives.

\subsection{Outline of the paper}

In this paper we suggest yet another microscopic mechanism allowing implementation of the coherent coupling to a biexciton and control over two-body interactions in a system of identical electromagnetic bosons. Our coherent link is due to the natural splitting of the radiative exciton doublet by the electron-hole exchange interaction. This genuinely excitonic mechanism is ubiquitous for a broad variety of 2D semiconductors including traditional quantum wells (QW's), atomically thin layers of transition metal dichalcogenides (TMD's) and emerging perovskite-based heterostructures. The coupling can be controlled \textit{in-situ} by in-plane structural anisotropy due to, \textit{e.g.}, uniaxial strain. 

Tuning the energy of the "open" channel close to the biexciton by a transverse magnetic field results in characteristic unitary behaviour of the $s$-wave boson scattering amplitude. For a broad resonance, this provides fully controllable and increasingly large third-order non-linear susceptibility  [$\chi^{(3)}$] at ultra-low photon densities. In microcavities the splitting of the polariton modes and the relative position of the scattering channels can additionally be controlled by a concomitant optical anisotropy of the cavity and the exciton-photon detuning, respectively. 

Our estimates of the resonance width indicate that one may expect efficient squeezing of the emitted photons in moderately strained TMD samples. The high efficiency of squeezing at the scattering resonance has also been previously predicted for dipolar polaritons,\cite{Andreev2020} although the control over the resonance width and its position relative to the scattering threshold is expected to be limited in that case. A generic physical reason underlying strong squeezing in both cases is that a resonant reaction of the type \eqref{ResonantReaction} implements an \textit{analog} second-order non-linear susceptibility [$\chi^{(2)}$] for electromagnetic waves. Unlike the conventional $\chi^{(2)}$, \cite{Boyd2008} the reaction \eqref{ResonantReaction} does not require any specific symmetry of the medium with respect to the inversion because the molecule $\alpha_{\uparrow}\alpha_\downarrow$ corresponds to a \textit{pair} of polarization waves. To distinguish ensuing parametric processes from the second-harmonic generation and the parametric down-conversion,\cite{Boyd2008} we shall refer to the former as to "fusion" and "disintegration" of electromagnetic bosons, respectively. Fusion (or disintegration) assumes macroscopic population of bosonic modes and occur on the characteristic timescale $\mathcal{T}_\alpha$ [Eq. \eqref{SqueezeTime}].

For excitons, we additionally predict giant Feshbach molecules (quantum halos) which can be obtained from a biexciton via \textit{rapid adiabatic passage} across the resonance. Alternatively, one may think of controllable synthesis of spin-singlet biexcitons from the polarized excitons. Biexciton synthesis (dissociation) is achieved through the reaction of the type \eqref{ResonantReaction} for two partilces, where one has $\alpha=\alpha^\prime=X$ and $\sigma=\sigma^\prime=\uparrow$, \footnote{In the quantum chemistry sense the formula \eqref{ResonantReaction} should be understood as a superposition [Eq. \eqref{PsiX}].} and is accompanied by a buildup (dissolution) of entanglement in the polarization of the emitted photons. This should be contrasted with the conventional process $X_\uparrow+X_\downarrow\leftrightarrows X_\uparrow X_\downarrow$ (potentially offered by the shape resonance,\cite{Andreev2016} although with limited control), where the degree of entanglement remains constant.  

Our consideration opens a new direction in the application-oriented quest of strongly-correlated photons and the quantum chemistry of light.

\section{Theoretcial model}
\label{Section2}

\subsection{Single-boson Hamiltonians}
\label{SubSec_IIA}

We adopt the generic picture of a "bright" exciton as an evanescent electromagnetic wave propagating in the structure plane. \cite{Ivanov1998, de2001evanescent} For a free-standing monolayer (or a QW) there is rapid radiative decay of excitons with low momenta $\bm{k}$ inside the light cone \cite{Hanamura1988}. The radiative exciton lifetime $\tau_X$ may be enhanced by placing the semiconductor onto a substrate \cite{Fang2019} or into a microcavity \cite{Voronova2018}. The dual view of an exciton as a bound electron-hole excitation and an electromagnetic wave then extends to the whole range of in-plane momenta. The conventional exciton-polariton in this picture represents the ultimate case where the strong light-matter coupling within the light cone is achieved by placing the 2D semiconductor at an anti-node of the microcavity photon mode.\cite{Weishbuch1992}

In the course of an optical transition the photon spin $\hat {\bm{s}}$ is transferred to the band electron (hole) in-plane orbital motion in the conduction (valence) band \cite{OpticalOrientation}. The same orbital momentum is responsible for the Zeeman-like interaction of an exciton with a static magnetic field $\bm{B}$. The three components of an expectation value $< \hat{\bm{s}} >$ are the Stokes parameters encoding the photon polarization \cite{QED}. In an unperturbed crystal the bright exciton states with $s_z=+1$ and $s_z=-1$ (hereinafter denoted as $\ket{\uparrow}$ and $\ket{\downarrow}$, respectively) form a degenerate doublet at $\bm{k}=0$. Under the time reversal one has $\hat s_z\rightarrow -\hat s_z$, whereas $\hat s_{x,y}\rightarrow \hat s_{x,y}$. Since, on the other hand, $\bm{B}\rightarrow-\bm{B}$, it follows that the exciton spin $\hat {\bm{s}}$ may couple only to the transverse component of $\bm{B}$. In the electron-hole picture this constraint corresponds to the mere fact that the 2D band orbital momentum does not possess an in-plane component \cite{Sallen2012, Zeng2012, Srivastava2015}.

Experimental \cite{Langer1968, Akimoto1968, Gourdon1992} and theoretical \cite{Fedorov1970, Pikus1970, Pikus1994, Glazov2022} studies of the optical orientation and alignment of excitons have shown that an effective in-plane magnetic field $\bm{\Omega}_\alpha$ may be induced by uniaxial deformation of the host lattice. Microscopically, the strain acts on the exciton spin via the electron-hole exchange interaction \cite{Bir1970}. In the basis of the circularly polarized states $\ket{\uparrow}$ and $\ket{\downarrow}$ the Hamiltonian of an electromagnetic boson under the combined action of the transverse magnetic field and the in-plane strain reads
\begin{equation}
\label{ExcitonHamiltonian}
\hat H_\alpha=\frac{\hbar^2 \hat {\bm{p}}^2}{2m_\alpha}+\frac{\delta_\alpha(B)}{2}-\hbar\bm{\omega}_\alpha\cdot \hat{\bm{s}},
\end{equation}
where $\hat {\bm{p}}$ is the momentum operator, $m_\alpha$ is the boson effective mass,
\begin{equation}    
\bm{\omega}_\alpha=\Omega_{\alpha,x}\bm n_x+\Omega_{\alpha,y}\bm n_y+\frac{\mu_B\mathrm{g}_\alpha}{\hbar} B \bm{n}_z
\end{equation}
and the boson spin operator may be expressed as
\begin{equation}
\hat {\bm{s}}=\hat\sigma_x \bm{n}_x+\hat\sigma_y \bm{n}_y+\hat\sigma_z \bm{n}_z
\end{equation}
with $\hat\sigma_x$, $\hat\sigma_y$ and $\hat\sigma_z$ being the Pauli matrices. We assume that the semiconductor band structure is such that the radiative doublet is isolated from various satellite states (\textit{e.g.}, "dark" excitons).\footnote{Beyond the bosonic model (\ref{ExcitonHamiltonian}), the exciton spin relaxation may also occur via sequential fermionic spin-flips involving the dark exciton as an intermediate state.\cite{Amand2017} The short-range electron-hole exchange suppresses such processes. In TMD's these are further suppressed due to the valleys in the absence of an inversion centre.\cite{Zeng2012,Mak2012}} Virtual transitions of the electromagnetic bosons to such states then may be accounted for by background renormalization of the relevant quantities. \cite{Ekardt1976}   

The label $\alpha$ distinguishes between the exciton ($\alpha=X$) and the lower-polariton ($\alpha=L$) models. For excitons, the term $\delta_X (B)$ describes the diamagnetic shift \cite{Deveaud2015,Crooker2018} and $\mathrm{g}_X$ is the exciton $\mathrm{g}$-factor which governs the Zeeman splitting in the transverse magnetic field $B$. In TMD monolayers one has $\mathrm{g}_X\sim 4$.\cite{Srivastava2015, Crooker2018} For polaritons, we shall assume large vacuum Rabi splitting $\hbar\Omega_R(B)$ as compared to the detuning $\delta_{\pm}=\delta_0-\delta_X(B)/2\pm\mu_B\mathrm{g}_X B$ between the microcavity and the exciton modes, \textit{i.e.} $\hbar\Omega_R(B)\gg |\delta_{\pm}|$. All polariton parameters vary along the dispersion curve and approach their bare excitonic values at high momenta. At the bottom of the dispersion one has $\delta_L(B)=\delta_X(B)/2+\delta_0-\hbar\Omega_R(B)$ and $\mathrm{g}_L=X^2_0 \mathrm{g}_X$ with $X^2_0=1/2[1+(1+\hbar^2\Omega_R^2/\delta_\pm^2)^{-1/2}]\approx 1/2$ being the Hoppfield coefficient $X^2_{\bm{p}}$ evaluated at $\bm{p}=0$. \cite{Deveaud2015}

A relationship between the effective magnetic field $\bm{\Omega}_\alpha$ and the strain tensor $u_{ij}$ may be established by using a symmetry argument. The scalar product in the Hamiltonian \eqref{ExcitonHamiltonian} should be invariant under the point symmetries of a 2D crystal. Hence, $\bm{\Omega}_\alpha$ belongs to the same (irreducible) representation as the in-plane coordinates $x$ and $y$. It follows, that $\Omega_{\alpha,x}=\mathcal{B}_\alpha(u_{xx}-u_{yy})$ and $\Omega_{\alpha,y}=2\mathcal{B}_\alpha u_{xy}$.

Assuming uniaxial deformation along the direction which constitutes an angle $\theta$ with the $x$ axis, the effective magnetic field may be expressed as
\begin{equation}
\label{EffectiveField}
\bm{\Omega}_\alpha=-\mathcal{B}_\alpha u [\cos (2\theta)\bm n_x-\sin (2\theta) \bm n_y],
\end{equation}
where $u\ll 1$ is the relative extension of the layer along the chosen direction, and we have assumed that the extension is accompanied by compression along the orthogonal direction by the same (relative) amount. An example of such deformation is sketched in Fig. \ref{Fig2}. In general, for $u_{xx}\neq -u_{yy}$ (the area is changed), one would also have an overall shift of the exciton energy proportional to $u_{xx}+u_{yy}$. \cite{Glazov2022} 

The proportionality coefficient $\mathcal{B}_\alpha$ depends on the particular setting under consideration and is governed by the electron-hole exchange. One has $\mathcal{B}_\alpha=\mathcal{B}_\alpha^{(\mathrm{sr})}+\mathcal{B}_\alpha^{(\mathrm{lr})}$, where the first and the second term are due to the short- and long-range parts of the exchange interaction, respectively. For excitons in TMD's, one has $\hbar\mathcal{B}_X^{(\mathrm{sr})}$ in the range $10$ to $100$ meV. \cite{Glazov2022} By placing the TMD layer onto a substrate with the amplitude reflection coefficient $r_b$, one may also activate the long-range part and increase the above estimate by the amount \cite{Glazov2022}
\begin{equation}
\label{LongRangePart}
\mathcal{B}_X^{(\mathrm{lr})}=\tau^{-1}_X\frac{2\eta \mathrm{Im}(r_b)}{1+\mathrm{Re}(r_b)}
\end{equation}
with $\eta\sim 1-10$. For polaritons, the coefficient $\mathcal{B}_L$ may additionally include purely photonic contribution due to an in-plane anisotropy of the microcavity. \cite{kaliteevskii, Bleu2017}

It is worth to point out that the long-range exchange (and the longitudinal-transverse splitting of the cavity mode) also provides the linear (quadratic) in $k$ contributions to the exciton (polariton) effective magnetic field $\bm{\Omega}_{X(L)}$ and kinetic energy. Possible effects due to those contributions have been discussed in our previous work \cite{Andreev2022_1} and do not affect the conclusions that will be drawn here.

Finally, we notice that all of the above mechanisms do not require any specific symmetry of the medium with respect to the spatial inversion. \cite{Glazov2022} In the absence of an inversion centre, one may have additional contribution to the effective field $\bm\Omega_\alpha$ due to the piezoelectric effect. \cite{Peng2018}

\subsection{Two-body interactions in the absence of strain}
\label{SubSec_IIB}
 
The Coulomb forces between the electrons and holes enable two-body interactions between the excitons. Thus, exchange of the identical fermions results in binding of excitons into biexcitonic molecules. \cite{Moskalenko, Culik1966} By virtue of the optical selection rule discussed above, opposite orientations of the fermionic spins in a biexciton imply the singlet configuration for the associated photons. The orbital wave function of a molecule in the absence of strain ($\bm{\Omega}_X\equiv 0$) may be conveniently regarded as an $\varepsilon<0$ solution of the Schr\"odinger equation
\begin{equation}
\label{Biexciton}
\left [\frac{\hbar^2 \hat {\bm{k}}^2}{m_X}+V_{\uparrow\downarrow}^{(X)}(r)\right]\varphi (r)=\varepsilon \varphi(r),
\end{equation}
where $ \hat {\bm{k}}=(\hat {\bm{p}}_1-\hat {\bm{p}}_2)/2$ is the relative momentum and $V_{\uparrow\downarrow}^{(X)}(r)$ is a static axially symmetric potential describing interaction of two composite bosons. \cite{Hanamura1974, Ekardt1976, Ivanov1998} The potential $V_{\uparrow\downarrow}^{(X)}(r)$ accounts both for the direct and exchange scatterings of the identical fermions, and in the heavy-hole limit approaches the familiar Heitler-London shape with the characteristic dip at short distances. For our analytical estimates we shall use the Gaussian \textit{ansatz}
\begin{equation}
\label{Gaussian}
\varphi (r)=(a\sqrt{\pi})^{-1}e^{-r^2/2a^2}
\end{equation}
with $a\equiv\hbar/\sqrt{m_X |\varepsilon|}$. We assume $a\sim R_{\uparrow\downarrow}^{(X)}$ with $R_{\uparrow\downarrow}^{(X)}$ being the microscopic range of the potential, \textit{i. e.}, our biexciton is a tightly-bound (deep) state. Thus, for TMD monolayers one has $a$ and $|\varepsilon|$ on the order of few nanometers and tens of meV, respectively.\cite{XXinTMD1, XXinTMD3, Hao2017} Studies of an analogous dipositronium problem suggest that excited states of a biexciton may actually not exist due to a very diffuse structure of a four-particle complex consisting of electrons and holes with equal masses \cite{Varandas2015}.

The excitons with parallel spins interact via the short-range repulsive potentials $V_{\sigma\sigma}^{(X)}(r)$.\footnote{Repulsive vs attractive characters of the potentials $V_{\uparrow\uparrow}^{(X)}(r)$, $V_{\downarrow\downarrow}^{(X)}(r)$ and $V_{\uparrow\downarrow}^{(X)}(r)$, respectively, have been confirmed by numerous experiments on dense ensembles, where the short-range correlations produce observable shifts of the exciton PL lines.\cite{Amand2017}} The corresponding low-energy $s$-wave scattering amplitudes may be written as\cite{Landau}
\begin{equation}
\label{Amplitudes}
\bar f_{\sigma\sigma^\prime}^{(X)}(k)=\frac{\pi}{\ln[k a_{\sigma\sigma^\prime}^{(X)}]-i\pi/2}
\end{equation}
with $a_{\sigma\sigma^\prime}^{(X)}$'s being the 2D scattering lengths. For the repulsive potentials $V_{\sigma\sigma}^{(X)}(r)$ one has $a_{\sigma\sigma}^{(X)}\sim R_{\sigma\sigma}^{(X)}$, where $R_{\sigma\sigma}^{(X)}$`s are the characteristic microscopic ranges. Such an estimate holds also for the attractive potential $V_{\uparrow\downarrow}^{(X)}(r)$ under our assumption of tight binding. 

Whether one deals with a bound state or a scattering event in the case of attraction depends on the statement of the problem: relaxation toward a bound state is beyond the elastic scattering considered here. In practice, the reaction $X_\uparrow+X_\downarrow\rightarrow X_\uparrow X_\downarrow$ due to the potential $V_{\uparrow\downarrow}^{(X)}(r)$ is an uncontrollable (stochastic) process, and the reverse process is of activation (Arrhenius) type. The situation may change for dipolar excitons, where the long-range dipolar repulsion introduces a potential barrier into $V_{\uparrow\downarrow}^{(X)}(r)$, which may transform the bound state into a shape resonance.\cite{Andreev2016} This may unlock a resonant interconversion process $X_\uparrow+X_\downarrow\leftrightarrows X_\uparrow X_\downarrow$, although the degree of control remains limited. We leave the long-range dipolar forces beyond the scope of the present paper as well.    

The effective interaction constants are related to the scattering amplitudes by $\bar g_{\sigma\sigma^\prime}^{(X)}=-2\hbar^2/m_X \bar f_{\sigma\sigma^\prime}^{(X)}(k_c)$, with $k_c$ being the characteristic momentum scale.\footnote{Here, one should put $k_c=\sqrt{2m_X k_B T}/\hbar$ for a classical gas of excitons at $T>T_\mathrm{BKT}$ and $k_c=\sqrt{2m_X \mu}/\hbar\sim \sqrt{n}$ for a quasi-condensate characterized by a positive chemical potential $\mu$ at $T<T_\mathrm{BKT}$, where $T_\mathrm{BKT}$ is the Berezinskii-Kosterlitz-Thouless transition temperature.\cite{Utesov} Note also the difference between a genuine binary mixture and our spin-$1$ system of electromagnetic bosons, where the odd-wave inter-species scattering is forbidden \cite{Andreev2022_1}} For a MoS$_2$ monolayer one may deduce from Eq. \eqref{Amplitudes} (omitting the imaginary part) typical values of  $\bar g_{\sigma\sigma^\prime}^{(X)}$'s on the order of few tenths of $\mu$eV$\times \mu$m$^2$. In the relevant limit 
\begin{equation}
\label{LowkLimit}
\sqrt{m_X/\hbar \tau_X}\ll k_c\ll 1/R_{\sigma\sigma^\prime}^{(X)}
\end{equation}
the three quantities $\bar g_{\sigma\sigma^\prime}^{(X)}$ together with the binding energy $|\varepsilon |$ provide complete description of the bare two-body interactions in a dilute exciton gas. Virtual transmutations of the bright excitons into remote (in the energy or momentum) satellite states may be regarded as background renormalization of these quantities \cite{Ekardt1976}.

For polaritons, we shall assume the Rabi splitting $\hbar\Omega_R$ being comparable to the biexciton binding energy $|\varepsilon|$. The polariton-polariton interactions at $\bm{\Omega}_L\equiv 0$ have been a subject of the on-going theoretical \cite{Byrnes2014, Bleu2020, Hu2022} and experimental \cite{Delteil2019, Estrecho2019, Stepanov2021, Snoke2023} studies. In most cases the projected potentials $V_{\sigma\sigma^\prime}^{(L)}(\bm{k}^\prime-\bm{k})\equiv X^2_{\bm{k}^\prime}X^2_{\bm{k}}V_{\sigma\sigma^\prime}^{(X)}(\bm{k}^\prime-\bm{k})$ may be approximated by some phenomenological constants $g_{\sigma\sigma^\prime}^{(L)}$ on the order of the corresponding effective interaction constants for excitons. 

The only exception is the situation where the polariton "$\uparrow\downarrow$" scattering continuum overlaps the biexciton level, \textit{i.e.},  
\begin{equation}
\label{PolaritonDetuning}
\delta_L (B)-\delta_X(B) < \varepsilon. 
\end{equation}
By using a toy model with a separable force \textit{in lieu} of the microscopic potential $V_{\uparrow\downarrow}^{(X)}(r)$ we obtain that in this limit the polariton scattering amplitude $\bar f_{\uparrow\downarrow}^{(L)}(k)$ is given by Eq. \eqref{Amplitudes} with $X\rightarrow L$ and 
\begin{equation}
\label{SingletPolaritonScattLength}
a_{\uparrow\downarrow}^{(L)}=\sqrt{\frac{m_X}{m_L}} e^{\tfrac{m_X}{m_L}\nu}a
\end{equation}
where  $\nu\propto[\varepsilon +\delta_X(B)-\delta_L (B)]/|\varepsilon|$. The corresponding effective interaction reads $g_{\uparrow\downarrow}^{(L)}=-2\hbar^2/m_L \bar f_{\uparrow\downarrow}^{(L)}(k)$. Since typically $m_L\ll m_X$ and, by assumption \eqref{PolaritonDetuning}, one has $\nu>0$, we deal with the scattering off a weakly-bound state. In 3D such scattering would produce increasingly strong repulsion. Here, in contrast, the scattering amplitude $\bar f_{\uparrow\downarrow}^{(L)}(k)$ vanishes.

The weakly-bound state occurs under the condition \eqref{PolaritonDetuning} as a pole of the scattering amplitude $\bar f_{\uparrow\downarrow}^{(L)}(k)$ at $k=i/a_{\uparrow\downarrow}^{(L)}$ and represents a hypothetical polariton molecule $L_\uparrow L_\downarrow$.\cite{LaRocca1998} A more involved study\cite{Ivanov1998} has shown, however, that instead of the reaction $L_\uparrow+L_\downarrow\leftrightarrows L_\uparrow L_\downarrow$, one would rather have an irreversible process $X_\uparrow X_\downarrow\rightarrow L_\uparrow+L_\downarrow $, which may be interpreted as autodissociation of a biexciton $X_\uparrow X_\downarrow$.\cite{Maruani1981} Such non-stationary biexciton in a microcavity has been referred to as "bipolariton".\cite{Ivanov1998}   

The effective interactions $g_{\sigma\sigma^\prime}^{(\alpha)}$ provide the third-order non-linear susceptibilities [$\chi^{(3)}$] for electromagnetic waves. \cite{Schumacher2007} As an example, two-body polariton scattering has been shown to produce degenerate parametric amplification of a signal upon coherent pump at specific angle fulfilling momentum conservation (\textit{i. e.}, the phase-matching condition) on the polariton dispersion.\cite{Savvidis2000}
  
\section{Results and discussion}

Consider a system of two electromagnetic bosons (excitons or polaritons). There are four basis states $\ket{\uparrow\uparrow}$, $\ket{\uparrow\downarrow}$, $\ket{\downarrow\uparrow}$, $\ket{\downarrow\downarrow}$ whose linear combinations realize the states $S_z=+2,0,-2$ of the total spin $S_z=s_{z,1}+s_{z,2}$. We notice that since $\bm{\Omega}_\alpha$ (the label $\alpha$ standing for $X$ or $L$, respectively) lies in the structure plane, the sum $\bm{\Omega}_\alpha\cdot \hat{\bm{s}}_1+\bm{\Omega}_\alpha\cdot \hat{\bm{s}}_2$ does not commute with $S_z^2$. Hence, the uniaxial deformation may change the spin state of a pair by flipping the spin of either of the two quasiparticles.

This observation implies that, in general, the wave function of the relative motion $\Psi^{(\alpha)}(r)$ would be a superposition of the corresponding three spin states, \textit{i.e.}
\begin{equation}
\Psi^{(\alpha)}(r)=\ket{\uparrow\uparrow}\psi_{\uparrow\uparrow}^{(\alpha)}(r)+\ket{\downarrow\downarrow}\psi_{\downarrow\downarrow}^{(\alpha)}(r)+\frac{1}{\sqrt{2}}(\ket{\uparrow\downarrow}+\ket{\downarrow\uparrow})\phi_{\uparrow\downarrow}^{(\alpha)}(r).
\end{equation}
By introducing the spin-flip matrix element $\Omega^{(\alpha)}_{\uparrow\downarrow}\equiv\hbar\bm{\Omega}_\alpha\cdot \braket{\uparrow|\hat {\bm{s}}|\downarrow} $ we may write the corresponding system of coupled Schr\"odinger equations in the form
\begin{equation}
\label{MultiChannel}
\left(-\frac{\hbar^2}{m_\alpha}\bm{\nabla}_{\bm{r}}^2+\hat V_\alpha\right)\Psi^{(\alpha)}(r)=E\Psi^{(\alpha)}(r)
\end{equation}
where
\begin{widetext}
\begin{equation}
\hat V_\alpha\equiv\left (\begin{array}{ccc}
V_{\uparrow\uparrow}^{(\alpha)}(r)+\delta_\alpha (B)-2\mu_B \mathrm{g}_\alpha B & 0 & -\sqrt{2}\Omega^{(\alpha)}_{\uparrow\downarrow}\\
0 &V_{\downarrow\downarrow}^{(\alpha)}(r)+\delta_\alpha (B)+2\mu_B \mathrm{g}_\alpha B & -\sqrt{2}\Omega^{(\alpha)}_{\downarrow\uparrow}\\
 -\sqrt{2}\Omega^{(\alpha)}_{\downarrow\uparrow} &  -\sqrt{2}\Omega^{(\alpha)}_{\uparrow\downarrow} &V_{\uparrow\downarrow}^{(\alpha)}(r)+\delta_\alpha(B)
 \end{array}\right).
 \end{equation}
 \end{widetext}
We are interested in a stationary scattering solution of this system near the lowest energy scattering threshold, \textit{i.e.} $E\rightarrow E^{(\alpha)}_{\bm{k}}\equiv\hbar^2 k^2/m_\alpha+\delta_\alpha-2\mu_B \mathrm{g}_\alpha B$ and $k\rightarrow 0$. The scattering channels with $S_z=0,-2$ are thus energetically closed. At $\bm{\Omega}_\alpha\equiv 0$ the solution is just a plane wave weakly distorted by the background potential:
\begin{equation}
\label{OpenChannelSolution}
\psi_{\uparrow\uparrow,\bm{k}}^{(\alpha)}(r)=\frac{1}{2\pi}\left (e^{i\bm{k}\bm{r}}+\frac{\bar f_{\uparrow\uparrow}^{(\alpha)}(k)}{\sqrt{-2\pi i k r}}e^{ikr}\right),
\end{equation}
where we have retained only the leading $s$-wave contribution into the (even) scattering amplitude of identical bosons. For excitons $f_{\uparrow\uparrow}^{(X)}(k)$ is given by Eq. \eqref{Amplitudes} and similar expressions may be written for polaritons (see Annexe \ref{AnnexeA}).

Provided the two conditions \textbf{(i)} $|\Delta_\alpha |\ll 2\mu_B \mathrm{g}_\alpha B$ and \textbf{(ii)} $| \Omega^{(\alpha)}_{\uparrow\downarrow} |\ll  \mu_B \mathrm{g}_\alpha B$ are fulfilled, the coherent coupling of the open ($S_z=+2$)  channel to the biexciton ($S_z=0$) due to the finite $\bm{\Omega}_\alpha$ results in resonant modification of the scattering amplitude. Here 
\begin{equation}
\bar\Delta_\alpha\equiv\varepsilon+\delta_X(B)+2\mu_B\mathrm{g}_\alpha B-\delta_\alpha(B)
\end{equation}
is the bare detuning between the open channel and the biexciton. The condition \textbf{(i)} suppresses the contribution due to the continuum of the $S_z=0$ channel. For excitons this condition reduces to $| 2\mu_B\mathrm{g}_X B+\varepsilon |\ll  \mu_B\mathrm{g}_L B$, whereas for polaritons it reads 
$| 2\mu_B\mathrm{g}_L B+\varepsilon-\delta_L (B)+\delta_X(B) |\ll  2\mu_B\mathrm{g}_L B$. The latter inequality requires $\delta_L (B)-\delta_X(B) >\varepsilon$ and, therefore, also excludes uncontrollable broadening of the biexciton level due to autodissociation into the "$\uparrow\downarrow$" polariton continuum.\cite{Ivanov1998} The condition \textbf{(ii)} suppresses the contribution of the  $S_z=-2$ channel.  
  
Under the assumptions \textbf{(i)} and \textbf{(ii)} the stationary scattering solution for the open channel retains the form of Eq. \eqref{OpenChannelSolution}, where one should substitute the bare $s$-wave scattering amplitude $\bar f_{\uparrow\uparrow}^{(\alpha)}(k)$ by $f_{\uparrow\uparrow}^{(\alpha)}(k)=\bar f_{\uparrow\uparrow}^{(\alpha)}(k)+f_{\uparrow\uparrow, \mathrm{res}}^{(\alpha)}(k)$ with
\begin{widetext} 
\begin{equation}
\label{ResAmplitude}
f_{\uparrow\uparrow, \mathrm{res}}^{(\alpha)}(k)=\frac{\pi \wp_\alpha}{(-\hbar^2 k^2/m_\alpha+\Delta_\alpha)/2\beta_\alpha+\ln(k a_\alpha)-i\pi /2}.
\end{equation}
\end{widetext}
Here
\begin{equation}
\label{beta}
\beta_\alpha\equiv 2m_\alpha a^2\Omega_\alpha^2/\hbar^2
\end{equation}
is the resonance width and $a$ is the biexciton radius as defined by the Gaussian \textit{ansatz} (\ref{Gaussian}). The factor $\wp_\alpha$ accounts for the distortion of the continuum in the open channel by the background potential $V_{\uparrow\uparrow}^{(\alpha)}(r)$.

The resonance is defined by
\begin{equation*} 
\Delta_\alpha=\Delta_\alpha^{(\mathrm{res})}\equiv-2\beta_\alpha \ln(k a_\alpha),
\end{equation*}
 and it is shifted toward lower energies with respect to the bare biexciton level $\varepsilon$. Such shift is a usual second-order coupling effect and may qualitatively be interpreted as polaronic "weighting" of the closed-channel biexciton. In the formal limit $k\rightarrow 0$ the scattering amplitude \eqref{ResAmplitude} at the resonance takes the value $f_{\uparrow\uparrow, \mathrm{res}}^{(\alpha)}(k\rightarrow 0)=2\wp_\alpha i/[1-2\mathrm{Arg}(a_\alpha)]$ which depends only on the background interaction.

In an ensemble of electromagnetic bosons, the resonant scattering amplitude \eqref{ResAmplitude} provides an additional contribution to the effective interaction $g_{\uparrow\uparrow}^{(\alpha)}$, and it differs dramatically from the corresponding background part discussed in Subsection \ref{SubSec_IIB}. At finite boson density $n$ one should distinguish between a narrow and a broad resonance, defined by the relations $\beta_\alpha\ll \hbar^2 k_c^2/m_\alpha$ and $\beta_\alpha\gg \hbar^2 k_c^2/m_\alpha$, respectively. Here $k_c$ is the characteristic value of the boson momentum in the ensemble: in a mean-field regime it is on the order of $k_c\sim \sqrt{n}$. For a narrow resonance the scattering amplitude at $\Delta_\alpha^{(\mathrm{res})}$ saturates at $f_{\uparrow\uparrow, \mathrm{res}}^{(\alpha)}(k_c)\sim -m_\alpha \wp_\alpha\beta_\alpha /\hbar^2 n\ll 1$, which yields the effective interaction $g_{\uparrow\uparrow, \mathrm{res}}^{(\alpha)}\sim  \wp_\alpha\beta_\alpha/n$ and a (positive) mean-field shift of the emitted photon energy $\sim \wp_\alpha\beta_\alpha $, which is much less than the corresponding contribution from the background scattering. For a broad resonance, one obtains the unitary value $f_{\uparrow\uparrow, \mathrm{res}}^{(\alpha)}(k_c)=2\wp_\alpha i/[1-2\mathrm{Arg}(a_\alpha)]$, so that the relation $g_{\uparrow\uparrow,\mathrm{res}}^{(\alpha)}=-2\hbar^2/m_\alpha f_{\uparrow\uparrow, \mathrm{res}}^{(\alpha)}(k_c)$ does not apply. A proper way to derive the resonant part of effective interaction between the bosons in this case will be outlined below.

In order to extend the scope of our discussion beyond the two-body picture presented above, we introduce a second-quantized many-body Hamiltonian\cite{Andreev2022_1}
\begin{equation}
\label{SecondHamiltonian}
\begin{split}
&\hat H_\alpha=\sum_{\bm k}\left[\frac{\hbar^2 k^2}{2m_\alpha}+\frac{\delta_\alpha(B)}{2}-\mu_B \mathrm{g}_\alpha B\right]\hat \alpha^\dagger_{\uparrow,\bm k}\hat \alpha_{\uparrow,\bm k}\\
&+\sum_{\bm K}\left[\frac{\hbar^2 K^2}{4m_X}+\varepsilon+\delta_X(B)\right] \hat C_{\uparrow\downarrow,\bm K}^\dagger  \hat C_{\uparrow\downarrow,\bm K}\\
&+\frac{\bar g_{\uparrow\uparrow}^{(\alpha)}}{2S}\sum_{\bm p,\bm p^\prime,\bm q}\hat \alpha_{\uparrow,\bm p^\prime+\bm q}^\dagger \hat \alpha_{\uparrow,\bm p-\bm q}^\dagger \hat \alpha_{\uparrow,\bm p} \hat \alpha_{\uparrow,\bm p^\prime}\\
&-\Omega^{(\alpha)}_{\downarrow\uparrow}\sum_{\bm k}\phi(\bm k)\hat C_{\uparrow\downarrow,\bm K}\hat \alpha_{\uparrow,-\bm k+\bm K/2}^\dagger \hat \alpha_{\uparrow,\bm k+\bm K/2}^\dagger+\mathrm{H.c.},
\end{split}
\end{equation} 
that would yield Eq. \eqref{MultiChannel} and the result \eqref{ResAmplitude} in the particular limit of two bosons in vacuum. The operator $\hat C_{\uparrow\downarrow,\bm K}$ describes a biexciton with the center-of-mass wave vector $\bm K$, $\hat \alpha_{\sigma,\bm p}$'s stand for single bosons and $\phi(\bm k)$ is the Fourier transform of Eq. \eqref{Gaussian}. The coherent interconversion term in the last line of Eq. \eqref{SecondHamiltonian} implements a reaction
\begin{equation}
\label{ResReaction}
\alpha_\uparrow+\alpha_\uparrow\leftrightarrows X_\uparrow X_\downarrow
\end{equation}
and has the form of $\chi^{(2)}$ non-linearity in quantum optics.\cite{Scully} Remarkably, such \textit{analog} $\chi^{(2)}$ does not require the absence of spatial inversion symmetry: the biexciton represents a \textit{pair} of electromagnetic waves and the matrix element $\Omega^{(\alpha)}_{\downarrow\uparrow}$ is perfectly allowed in centrosymmetric crystals (see Subsection \ref{SubSec_IIA}).

The reaction \eqref{ResReaction} is expected to be particularly efficient in the broad resonance limit $\beta_\alpha\gg \hbar^2 k_c^2/m_\alpha$. In this limit, the biexciton operator $\hat C_{\uparrow\downarrow,\bm K}$ may be shown to play the role of an \textit{auxiliary} field which describes onset of pair correlations between the bosons $\alpha_\uparrow$.\cite{Andreev2020} One may then adiabatically eliminate $\hat C_{\uparrow\downarrow,\bm K}$ from the Hamiltonian \eqref{SecondHamiltonian} to obtain \cite{Andreev2020}
\begin{equation}
\label{ReducedAmplitude}
g_{\uparrow\uparrow}^{(\alpha)}=\bar g_{\uparrow\uparrow}^{(\alpha)}-\frac{\hbar^2}{ m_\alpha}\frac{2\pi \beta_\alpha}{\Delta_\alpha}.
\end{equation}
The effective interaction $g_{\uparrow\uparrow}^{(\alpha)}$ provides increasingly large and tuneable $\chi^{(3)}$.\footnote{The average kinetic energy of the relative motion in the denominator of Eq. \eqref{ReducedAmplitude} cancels in the case of massive, parabolic dispersion assumed here.} Simultaneous presence of efficient second-order $[\chi^{(2)}]$ and divergent third-order $[\chi^{(3)}]$ non-linearities reflects a dual nature of the solution of the many-body problem \eqref{SecondHamiltonian} at unitarity.\cite{Pricoupenko2004}

A natural manifestation of $\chi^{(2)}$ would be second-harmonic generation and the corresponding reverse process known as parametric down conversion.\cite{Boyd2008} Both of these processes exhibit coherent intensity transfers between electromagnetic waves. In the  quantum chemistry parlance, this can be seen as the Bose stimulation of the reaction \eqref{ResReaction} in a macroscopic ensemble.\cite{Heinzen2000, Moore2002, Zhang2023} In our case, we may naturally think of such processes as of controllable \textit{fusion} and \textit{disintegration} of electromagnetic bosons, respectively. Following the lines of our earlier work on dipolar polaritons,\cite{Andreev2020} one may deduce from Eq. \eqref{SecondHamiltonian} that at unitarity ($\Delta_\alpha=0$) the fusion of bosons $\alpha_\uparrow$ occurs on the characteristic timescale $\mathcal{T}_\alpha=\sqrt{m_\alpha/2\pi \beta_\alpha n}$, which upon substitution of $\beta_\alpha$ from Eq. \eqref{beta} may be recast as
\begin{equation}
\label{SqueezeTime}
\mathcal{T}_\alpha=(4\pi na^2)^{-1/2}\Omega_\alpha^{-1}.
\end{equation}
Simultaneously, quantum squeezing of electromagnetic waves is observed on this timescale.\cite{Andreev2020} The squeezing is indicative of strong correlations near resonance. The result \eqref{ResAmplitude} offers a way to explore controllable fusion and strong correlations in the ultimate limit of two bosons in vacuum. We delve deeper into this intriguing scenario in the subsequent section.

Naively, one would expect that a broad resonance would be beneficial for fusion and squeezing at low densities. In practice, however, one should compare the timescale $\mathcal{T}_\alpha$ with the boson radiative lifetime $\tau_\alpha$. Faster radiative decay may provide stronger coupling $|\Omega^{(\alpha)}_{\uparrow\downarrow}|$ (see Subsection \ref{SubSec_IIA}), but a photon may escape the crystal before the reaction \eqref{ResReaction} takes place. Also compatibility of a broad resonance with the condition $\textbf{(ii)}$ should be discussed with care.  

Although the above equations in their generic forms apply both for excitons ($\alpha=X$) and lower polaritons ($\alpha=L$), the actual results as defined by the respective parameters are very different. Below we discuss these two distinct cases separately and in greater detail.
 
\subsection{Excitons ($\alpha=X$)}
\label{SubSec_IIIA} 

We obtain $\Delta_X=\bar\Delta_X=\varepsilon+2\mu_B\mathrm{g}_X B$ and 
\begin{equation*}
a_X= a e^{\gamma/2},
\end{equation*}
so that the exciton scattering amplitude \eqref{ResAmplitude} takes the genuine $s$-wave resonant form. This implies the resonant pairing phenomenology, echoing the physics of a shape resonance suggested previously for dipolar excitons \cite{Andreev2016} and polaritons \cite{Andreev2020}. Treating the colliding excitons as impenetrable disks of radius $R_{\uparrow\uparrow}^{(X)}$ and assuming $R_{\uparrow\uparrow}^{(X)}\ll a$ yields the estimate (see Annexe \ref{AnnexeC})
\begin{equation}
\label{Factor}
\wp_X\sim \left (\frac{\ln{R_{\uparrow\uparrow}^{(X)}/a}}{\ln{k R_{\uparrow\uparrow}^{(X)}}}\right)^2.
\end{equation}
Despite the apparent crudeness of such approximation for $V_{\uparrow\uparrow}^{(X)}(r)$, the result (\ref{Factor}) correctly captures the 2D kinematics: the excitons tend to avoid each other at $k\rightarrow 0$, which results in suppression of the coherent coupling to the biexciton.

The condition $\textbf{(ii)}$ implies  $|\varepsilon|\gg \beta_X$. This inequality may also be recast in a more conventional form as $(\hbar\Omega_X/\varepsilon)^2\ll 1$. It is instructive to consider compatibility of this condition with the narrow ($\beta_X\ll \hbar^2k_c^2/m_X$)  \textit{vs} broad ($\beta_X\gg \hbar^2k_c^2/m_X$) resonance regimes. By using the definition \eqref{beta} and assuming $k_c\sim\sqrt{n}$, these regimes may be conveniently recast as $(\hbar\Omega_X/\varepsilon)^2\ll na^2$ and $(\hbar\Omega_X/\varepsilon)^2\gg na^2$, respectively. Since, generically, $na^2\lesssim 1$, one can see, that a narrow resonance is always compatible with $\textbf{(ii)}$, whereas a broad resonance would additionally require $na^2\ll 1$.     

\begin{figure}[t]
\centering
\includegraphics[width=0.8\columnwidth]{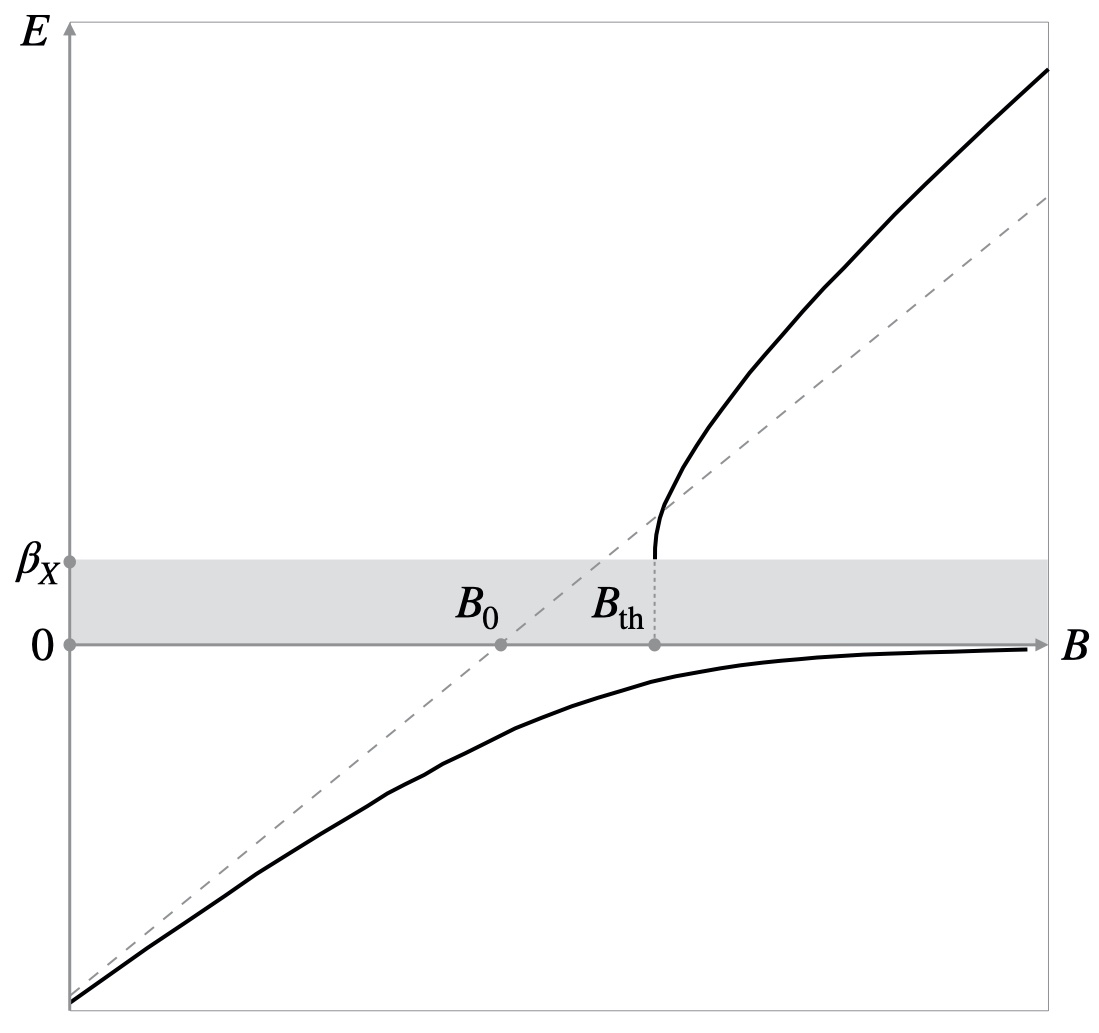}
\caption{The poles $E$ of the exciton $s$-wave scattering amplitude $f_{\uparrow\uparrow, \mathrm{res}}^{(X)}(k)$ as functions of the magnetic field $B$. The full equation for the pair polarization bubble has been used (see Annexe \ref{AnnexeB}). At $B>B_\mathrm{th}$ the weakly-bound state coexists with a resonance at $E>0$  (only the real part is shown). The dashed line indicates the asymptote $E(B)=2\mu_B\mathrm g_X(B-B_0)$. At $B\gg B_\mathrm{th}$ the wave function of the weakly-bound state takes the universal form of the quantum halo.}
\label{Fig1}
\end{figure}

Being considered as a complex function of the energy $E=\hbar^2 k^2/m_X$ the resonant part of the exciton scattering amplitude (\ref{ResAmplitude}) has poles defined by the equation $E=\bar\Delta_X-\beta_X\ln(|\varepsilon|e^{-\gamma}/E)-i\pi$. At negative detuning $\bar\Delta_X<0$ there is a single pole that asymptotically approaches the straight line $E=\bar\Delta_X$ as $\bar\Delta_X\rightarrow-\infty$. By writing $\bar\Delta_X(B)=2\mu_B\mathrm{g}_X (B-B_0)$ and considering the energy $E$ as a function of $B$, we may let $E(0)\approx \varepsilon$. As the magnetic field $B$ exceeds the threshold value
\begin{equation}
B_\mathrm{th}=B_0+\beta_X (2\mu_B\mathrm{g}_X)^{-1}[1+\ln(|\varepsilon|e^{-\gamma}/\beta_X)],
\end{equation}
a second pole emerges at positive energies (see Fig. \ref{Fig1}). This pole is a resonance characterized by the decay rate $\propto \beta_X$. Interestingly, the resonance coexists with the weakly-bound state
\begin{equation}
\label{HaloEnergy}
E(B)=-|\varepsilon |e^{-\gamma}e^{-2\mu_B \mathrm{g}_X B/\beta_X}
\end{equation}
at $B\gg B_\mathrm{th}$. This should be contrasted to the $s$-wave Feshbach scattering in 3D \cite{Gurarie2007} and scattering in higher partial-wave channels in 2D, \cite{Andreev2022_1,Andreev2022_2} where the bound state becomes substituted by the resonance.

\begin{figure}[t]
\centering
\includegraphics[width=1\columnwidth]{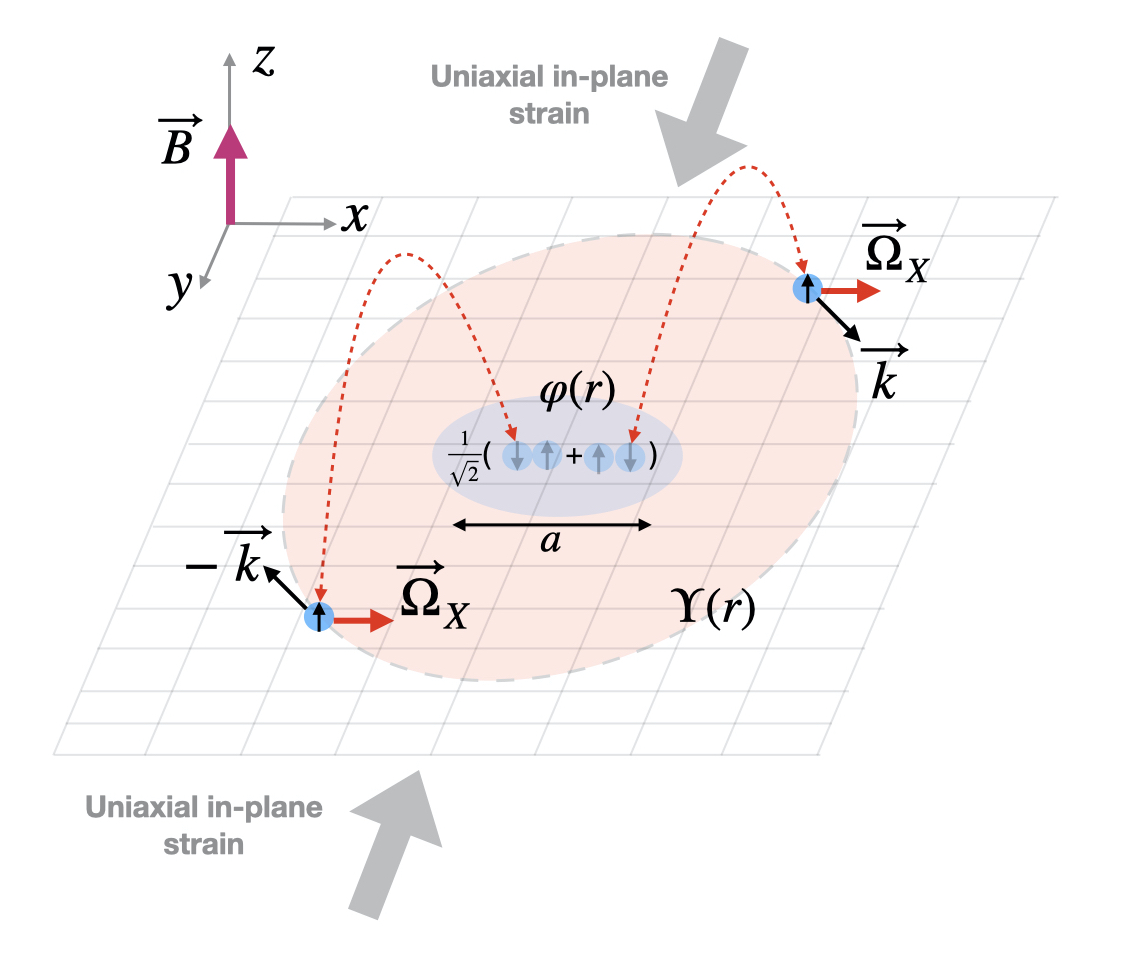}
\caption{Schematic view of a synthetic excitonic molecule [Eq. \eqref{PsiX}] obtained in a strained 2D semiconductor by sweeping the tranverse magnetic field $\bm{B}$ across the biexciton resonance. Uniaxial strain applied along the $y$ direction induces an effective magnetic field $\bm\Omega_X$ oriented along the $x$ axis. In contrast to $\bm{B}$, the field $\bm{\Omega}_X$ is even under time reversal and couples spin-polarized free-pair states $\ket{\bm{k},\uparrow}\ket{-\bm{k},\uparrow}$ to the spin-entangled biexciton state $\tfrac{1}{\sqrt{2}}(\ket{\uparrow\downarrow}+\ket{\downarrow\uparrow})$ with the spatial part $\varphi(r)$ (blue area) having microscopic size $a$. Such coupling (depicted by dashed red arrows) acts as a coherent Feshbach link which "dresses" the biexciton by the spin-polarized continuum, the latter forming a quantum halo $\Upsilon (r)$ which extends over the macroscopic area (pink).}
\label{Fig2}
\end{figure}

The wave function of the bound state can be written in the form
\begin{equation}
\label{PsiX}
\Psi(\bm{r})=\Upsilon(r)\ket{\uparrow\uparrow}+w\varphi (r)\frac{1}{\sqrt{2}}(\ket{\uparrow\downarrow}+\ket{\downarrow\uparrow}), 
\end{equation}
where $\varphi (r)$ is the tightly-bound core given by Eq. \eqref{Gaussian} and $\Upsilon(r)$ is the so-called \textit{quantum halo} which extends well beyond the microscopic range of the potential $V_{\uparrow\downarrow}^{(X)}(r)$ (Fig. \ref{Fig2}). By using Eq. \eqref{HaloEnergy} and the relation\cite{Andreev2022_1} $w^2=(2\mu_Bg)^{-1}\partial E/\partial B$, the relative weight of the core $w^2$ may be expressed as
\begin{equation}
\label{RelativeWeight} 
w^2=\frac{|\varepsilon|}{\beta_X}e^{-\gamma}e^{-2\mu_B \mathrm{g}_X B/\beta_X}.
\end{equation}
One can see that $w$ rapidly approaches zero as $B$ is tuned beyond the threshold $B_\mathrm{th}$, so that the contribution of the halo to the state (\ref{PsiX}) becomes dominant. In the same limit, the halo takes the form
\begin{equation}
\label{Halo}  
\Upsilon(r)=\frac{1}{\sqrt{\pi a_\Upsilon^2}}K_0(r/a_\Upsilon)
\end{equation} 
with $a_\Upsilon \equiv \hbar/\sqrt{m_X E(B)}\rightarrow\infty$ as $E(B)\rightarrow 0$.  Eq. \eqref{Halo} is universal in the sense that it does not depend on the details of the microscopic interaction potential. In the opposite limit $B\rightarrow 0$, the relative weight $w^2$ approaches unity: the bound state becomes the bare biexciton.

Let us discuss manifestation of the predicted phenomenology in optical experiments. The most basic quantity is the exciton oscillator strength $f_X$ which defines the exciton radiative decay $\Gamma_X$. The oscillator strength scales with the coherence area covered by the exciton c.o.m. translational motion, \cite{Elliott1957, Rashba1962} which for the weakly-bound state \eqref{Halo} may be expressed as  $f_\Upsilon=f_X |\Upsilon_{0}|^2$, where $\Upsilon_{0}$ is the $\bm{p}=0$ value of the Fourier transform $\Upsilon_{\bm{p}}=\sqrt{4\pi a_\Upsilon^2/S}/(1+p^2 a_\Upsilon^2)$ of Eq. \eqref{Halo}. The upper bound on $a_\Upsilon$ is defined by the condition \eqref{LowkLimit} and, ultimately - by the quantization area $S$ (the area of the 2D structure). The oscillator strength of an exciton constituting the halo approaches that of a free exciton (plane wave) as $a_\Upsilon\rightarrow\sqrt{S/4\pi}$. On the other hand, the oscillator strength of the halo is enhanced with respect to the bare biexciton by the factor $(a_\Upsilon/a)^2$. \footnote{This estimate does not account for the extrusion of the average boson momenta beyond the "light cone" due to the tight binding in real space. \cite{Citrin1994} However, this additional effect is of secondary importance here.}  Hence, the emergence of the halo should manifest itself in exponential growth and subsequent saturation of $\Gamma_\Upsilon$, as $B$ is swept across $B_\mathrm{th}$. The sweeping of the magnetic field in this case should be performed adiabatically starting from $B=0$, which would require sufficiently long lifetime $\tau_X$ of the bare exciton, \textit{i.e.} $\tau_X\gg \hbar/\beta_X$ with $\beta_X$ being the resonance width given by Eq. \eqref{beta}.

Alternatively, one may think of controllable assembly (synthesis) of a biexciton from a polarized continuum, akin to the Feshbach ramp in ultra-cold atoms. \cite{Chin2010}
Interestingly, the three-body recombination, which is known to be a major side effect for synthesis of atomic molecules\cite{Chin2010}, is absent for excitons. There are three factors that preclude such an inelastic process. First, as we have pointed out above, the exciton halo is formed when the "open" channel is still energetically \textit{below} the resonance. Second, the probability for three excitons to approach each other is suppressed by the Pauli principle for the constituent electrons and holes. Deep bound states of three excitons do not exist (although one cannot exclude trimers of macroscopically large radius \cite{Aleiner2017}). Finally, the level $\varepsilon$ is likely to be a unique bound state of the potential $V_{\uparrow\downarrow}^{(X)}(r)$ (see Subsection \ref{SubSec_IIB}), and the polarized exciton continuum couples directly to this level. These factors hold promise for experimental implementation of synthetic molecules and their practical use, as we discuss below.                 

The polarization part of the wavefunction \eqref{PsiX} evolves from a product state $\ket{\uparrow\uparrow}$ to a maximally entangled state $\frac{1}{\sqrt{2}}(\ket{\uparrow\downarrow}+\ket{\downarrow\uparrow})$ as the magnetic field is tuned from $B\gg B_\mathrm{th}$ to $0$. To quantify the entanglement of the polarization, we evaluate the \textit{concurrence} \cite{Mintert2005, Wootters1998}  
\begin{equation}
\label{concurrence}
c[\hat\rho]\equiv \inf\limits_{\{p_i,\Phi_i\}}\sum_i p_i \bar c[\Phi_i],
\end{equation}
where the infimum is searched over all possible convex decompositions $\{p_i,\Phi_i\}$ of the mixed state
\begin{equation}
\label{PolDensMatrix}
\hat\rho\equiv\int \Psi(\bm{r})\Psi^{\dagger}(\bm{r})d\bm{r} =\sum_i p_i\ket{\Phi_i}\bra{\Phi_i}
\end{equation}
obtained by tracing out the spatial part of the wavefunction \eqref{Halo}. Here
\begin{equation}
\label{PureConcurrence}
\bar c[\Phi_i]\equiv|\bra{\Phi_i^\ast}\hat\sigma_y\otimes\hat\sigma_y\ket{\Phi_i}|
\end{equation} 
are concurrencies of the pure bipartite polarization states $\ket{\Phi_i}$. 

Although the density matrix $\hat\rho$ depends parametrically on the spatial overlap of the core $\varphi(r)$ and the halo $\Upsilon(r)$, we find that the concurrence $c[\hat\rho]$ does not depend on this parameter and, in fact, is identical to that of a pure polarization state
\begin{equation}
\ket{\Phi}\equiv \sqrt{1-w^2}\ket{\uparrow\uparrow}+w\frac{1}{\sqrt{2}}(\ket{\uparrow\downarrow}+\ket{\downarrow\uparrow}),
\end{equation}
which formally may be obtained by dropping the spatial components in Eq. \eqref{PsiX}. Namely, we find
\begin{equation}
\label{ConcurrenceRelation}
c[\hat\rho]=\bar c[\Phi]=w^2,
\end{equation}
where $\bar c[\Phi]$ is calculated by using Eq. \eqref{PureConcurrence} and the relative weight $w^2$ depends on the applied magnetic field according to Eq. \eqref{RelativeWeight}. We provide detailed derivation of the relation \eqref{ConcurrenceRelation} in Annexe \ref{AnnexeD}. The auxiliary state $\ket{\Phi}$ then allows one to develop a better feel for the entanglement measure \eqref{concurrence} by resorting to the alternative version of Eq. \eqref{PureConcurrence} for pure states \cite{Mintert2005}       
\begin{equation}
\bar c[\Phi]=\left |\sum\limits_{i=1}^4\braket{e_i|\Phi}^2\right|
\end{equation}
with $\ket{e_1}=1/\sqrt{2}(\ket{\uparrow\uparrow}+\ket{\downarrow\downarrow})$, $\ket{e_2}=i/\sqrt{2}(\ket{\uparrow\uparrow}-\ket{\downarrow\downarrow})$,
$\ket{e_3}=i/\sqrt{2}(\ket{\uparrow\downarrow}+\ket{\downarrow\uparrow})$ and $\ket{e_4}=1/\sqrt{2}(\ket{\uparrow\downarrow}-\ket{\downarrow\uparrow})$ being the Bell states. One can see that $\bar c[\Phi]$ reaches its maximum value when the vector $\ket{\Phi}$ converges (up to an unimportant phase factor) to the Bell sate $\ket{e_3}$.

According to Eq. \eqref{ConcurrenceRelation}, the concurrence $c[\hat\rho]$ varies monotonously from $1$ to $0$ as the magnetic field is swept in the range $B\in [0,+\infty)$. Hence, similarly to the radiative decay rate $\Gamma$ discussed above, the entanglement may follow adiabatically sufficiently rapid variation of $B$. The ability to prepare an arbitrarily entangled state via a \textit{rapid adiabatic passage} renders the composite state \eqref{PsiX} appealing for the quantum information processing. For instance, controllable generation of the Bell state is known to be a crucial stage in the quantum teleportation algorithm.\cite{Riebe2004} Detailed proposals of such kind are beyond the scope of the present work and will be given elsewhere. 

A qualitatively different scenario may be realized in a scattering experiment whereby pairs of spin-$\uparrow$ excitons undergo elastic collisions governed either by the scattering amplitude \eqref{ResAmplitude} or by the effective interaction \eqref{ReducedAmplitude}, depending on whether one deals with a narrow or a broad resonance, respectively. For example, a spin-polarized gas of excitons may be resonantly pumped at $B$ close to $B_\mathrm{th}$ and then probed in reflectance (or transmittance) by a weak pulse of \textit{the same} circular polarization. The formula  \eqref{ReducedAmplitude} then predicts increasingly large blueshift or redshift of the probe signal depending on whether the energy of colliding bosons is above or below the resonance, respectively. Alternatively, one may examine the spectral shifts of the exciton PL signal and spatial distribution of the PL intensity. Consistent description of such experiments can be carried out starting from the Hamiltonian \eqref{SecondHamiltonian} and proceeding along the lines of the many-body theory previously designed for dipolar excitons \cite{Andreev2016} and polaritons \cite{Andreev2020} interacting via a shape resonance in the microscopic potential $V_{\uparrow\downarrow}(r)$. An advantage of the present setting is a possibility to control independently the detuning $\bar\Delta_X$ and the width $\beta_X$ of the resonance.

Let us consider the explicit example of excitons in a monolayer of MoS$_2$. For a free-standing monolayer the matrix element $|\Omega_{\uparrow\downarrow}^{(X)}|$ would be defined through Eq. \eqref{EffectiveField} by the short-range part of the exchange interaction. Taking the upper estimate for the corresponding parameter\cite{Glazov2022} $\hbar\mathcal{B}_X^{(\mathrm{sr})}\sim 100$ meV, the relative extension $u\sim 0.01$ and the biexciton binding energy\cite{Hao2017} $|\varepsilon|\sim 20$ meV, we obtain the ratio $(\hbar\Omega_X/\varepsilon)^2\sim 10^{-3}$, so that the resonance is expected to be narrow at densities down to $na^2\sim 10^{-2}$. An intermediate regime between a narrow and a broad resonance at such low densities could potentially be promising for observation of the halos. However, the resonance width $\beta_X\sim (\hbar\Omega_X)^2/|\varepsilon|\sim 0.05$ meV is much less than the exciton radiative decay rate $\Gamma_X$ (few meV), which may hinder rapid adiabatic preparation of such states as discussed above. Likewise, the lower bound on the characteristic time $\mathcal{T}_X$ (achieved at $na^2\sim 1$) is comparable to the exciton radiative lifetime $\tau_X\sim 0.1$ ps. This precludes fusion and squeezing.

The situation changes qualitatively in the presence of a substrate. The substrate characterized by the amplitude reflection coefficient $r_b=-e^{-i\delta\phi}$ with $\delta\phi\ll 1$ may increase the radiative lifetime $\tau_X$ by the huge factor $\delta\phi^{-2}$. Thus, in the experiment\cite{Fang2019} the factor $\delta\phi\sim 0.1$ has been achieved. The radiative decay rate $\Gamma_X$ then becomes smaller than the resonance width (as defined by the short-range exchange), and in magnetic fields on the order of few tens of Tesla\cite{Crooker2018} one should be able to realize controllable synthesis of biexcitons.

The upper bound on the molecular radius as defined by the condition \eqref{LowkLimit} would be $a_\Upsilon\lesssim  20 a$, which would require $na^2\lesssim 10^{-3}$. In the interval of temperatures $10^{-2}$ K $\ll T\lesssim$ $0.1$ K one would have a boltzmannian gas of molecules with their internal states being described by the pure states \eqref{Halo}. The polarization density matrix $\hat\rho$ [Eq. \eqref{PolDensMatrix}] in this case might be accessed experimentally by cross correlation polarization measurements in a photon coincidence setup analogous to that employed previously for QD's \cite{Stevenson2006} (see also Annexe \ref{AnnexeE}).

Furthermore, by using Eq. \eqref{LongRangePart} and Eq. \eqref{SqueezeTime}, the ratio of the fusion time to the exciton radiative lifetime may be estimated as
\begin{equation*}
\frac{\mathcal{T}_X}{\tau_X}\approx(4\pi na^2)^{-1/2}\frac{\delta\phi}{u}(\zeta/\delta\phi+2\eta)^{-1},
\end{equation*}
with $\zeta\sim 1-10$ being the dimensionless product of $\mathcal{B}_X^{(\mathrm{sr})}$ and the radiative lifetime in the absence of substrate, and the parameter $\eta$ coming from the long-range contribution to the exchange. This estimate shows that the ratio $\mathcal{T}_X/\tau_X\sim 0.1$ may be achieved at densities $na^2\sim 0.1$. The squeezing may be verified by examining the statistics of emitted photons with balanced homodyne detection \cite{Breitenbach1997}.

A different strategy might be applied to achieve tuneable $\chi^{(3)}$. One may take advantage of the long-range exchange in the presence of a substrate by choosing $r_b\approx e^{i\pi/2}$ and then trying to reach the regime $\beta_X\sim\Gamma_X$ by increasing the relative deformation $u$. Although $\mathcal{T}_X$ would substantially exceed $\tau_X$ in this case, thus precluding efficient $\chi^{(2)}$ processes, one nevertheless should be able to get the tuneable effective interaction \eqref{ReducedAmplitude} in a scattering experiment described above. 

\subsection{Lower polaritons ($\alpha=L$)}

\begin{figure}[t]
\centering
\includegraphics[width=1\columnwidth]{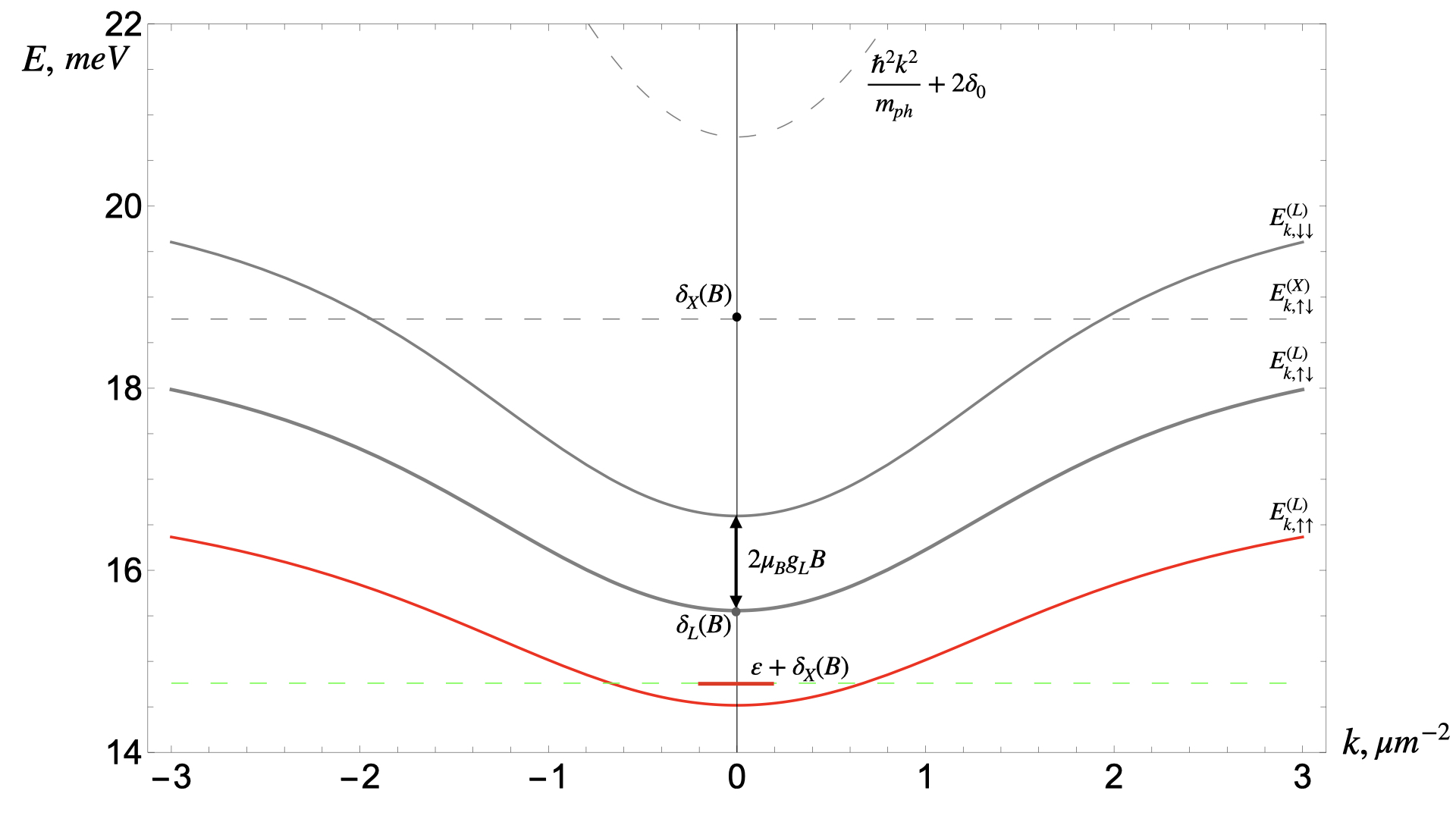}
\caption{Configuration of the lower polariton ($\alpha=L$) pair energy levels fulfilling the resonant approximation \textbf{(i)} used to derive the scattering amplitude Eq. \eqref{ResAmplitude}. The wave vector $k$ stands for the pair relative motion and we have put $K=0$ (the center-of-mass reference frame). Dashed parabola and horizontal line are the bare two-photon and two-exciton energies, respectively. Short horizontal bar indicates the biexciton energy level. The parameters typical for the GaAs/AlGaAs-based planar microcavities \cite{Deveaud2015} have been used: $m_X=0.05m_e$ ($m_e$ being the free electron mass), $\hbar\Omega_R=4$ meV, $|\varepsilon|=2$ meV, $g_X=1.35$, $\delta_X(B)=0.67\times B$ meV, and we take $B=14$ T. The two-body scattering process of interest is supposed to take place at the bottom of the lowest energy dispersion line denoted as $E_{k,\uparrow\uparrow}^{(L)}$.}
\label{Fig3}
\end{figure}

The configuration of two-particle energy levels fulfilling the condition \textbf{(i)} in this case is tentatively sketched in Fig. \ref{Fig3}. The scattering amplitude $f_{\uparrow\uparrow}^{(L)}(k)$ given by Eq. \eqref{ResAmplitude} refers to the scattering of two spin-$\uparrow$ polaritons at the bottom of the corresponding dispersion ($k\rightarrow 0$). In contrast to excitons, for polaritons we find $\Delta_L\equiv\bar\Delta_L+(m_X/m_L)\beta_L e\mathrm{Ei}(-1)$ and 
\begin{equation*}
a_L\equiv i\sqrt{\frac{m_X}{m_L}}a,
\end{equation*}
the latter quantity now being purely imaginary. The imaginary unit cancels with $i\pi/2$ in the denominator of Eq. \eqref{ResAmplitude} and thus eliminates the resonant pole structure. One can also show that for polaritons the effect of the background potential is negligible, so that $\wp_L=1$.

The absence of the pole structure in the scattering amplitude implies the absence of the polariton bound states and resonances. On the other hand, the scattering amplitude still possesses a resonant denominator, which should manifest itself in elastic collisions. To estimate the ratio of the resonance width $\beta_L$ to the characteristic kinetic energy of polaritons, we first notice that in a mean-field regime the latter is on the order of the corresponding energy for excitons, \textit{i. e.}, one has 
\begin{equation*}
\hbar^2 k_c^2/m_L\sim g_{\sigma\sigma^\prime}^{(L)} n\sim \hbar^2 n/m_X
\end{equation*}
(see Subsection \ref{SubSec_IIB} and Annexe \ref{AnnexeA}). We believe that such a mean-field estimate is pertinent to the vast majority of experimental cases, where coherent population of polaritons is generated at the bottom of their dispersion. \footnote{An alternative thermal estimate $\hbar^2 k_c^2/m_L\sim T_\mathrm{BKT}$, with $T_\mathrm{BKT}\sim \hbar^2 n/m_L$, may be orders of magnitude larger in the case of polaritons.} We may then estimate the relevant ratio as       
\begin{equation*}
\frac{\beta_L}{\hbar^2 k_c^2/m_L}\sim \frac{m_L}{m_X} \left(\frac{\Omega_L}{\varepsilon}\right)^2(na^2)^{-1},
\end{equation*}
and conclude that, since, typically, $\Omega_L\sim \Omega_X$, the polariton resonance formally remains narrow down to ultra-low densities. However, the energy shift of a probe light pulse in a hypothetical experiment of the type described in Subsection \ref{SubSec_IIIA} would be governed by the interaction constant 
\begin{equation}
\label{PolReducedAmplitude}
g_{\uparrow\uparrow, \mathrm{res}}^{(L)}\equiv-\frac{2\hbar^2}{m_L} f_{\uparrow\uparrow, \mathrm{res}}^{(L)}=\frac{\hbar^2}{m_L}\frac{2\pi \beta_L}{\hbar^2 k_c^2/m_L-\Delta_L},
\end{equation}
where we have omitted the logarithm in Eq. \eqref{ResAmplitude} assuming that $\Delta_L\sim \hbar^2 k_c^2/m_L\gg \beta_L$. The result \eqref{PolReducedAmplitude} is formally analogous to the effective interaction in the case of a broad resonance given by Eq. \eqref{ReducedAmplitude}, with the peculiar difference that now the position of the singularity would also depend on the density $n$ through $k_c$. The rescaled "resonance width" in the numerator would be comparable with $\beta_X$, and may be made larger than the kinetic energy and the polariton radiative decay rate. Likewise, the polariton fusion time $\mathcal{T}_L$ [Eq. \eqref{SqueezeTime}]  should be comparable with $\mathcal{T}_X$.

Magneto-optics of microcavities with TMD's being currently at its early stage of development, we consider polaritons in an epitaxial GaAs    
QW inside a planar GaAs/AlAs-based microcavity.\cite{Deveaud2015} An exemplary configuration of the lower polariton pair energy levels fulfilling the resonant approximation \textbf{(i)} is presented in Fig. \ref{Fig3}. One may expect to achieve linear polarization splittings $\hbar\Omega_L$ on the order of 1meV in such systems under moderate anisotropy.\cite{kaliteevskii} We then obtain $(\hbar\Omega_L)^2/|\varepsilon|\sim 1$ meV for the rescaled "resonance width". This indeed exceeds the polariton radiative decay rate $\Gamma_L\sim 0.1$ meV by an order of magnitude. The ratio $\mathcal{T}_L/\tau_L$ may also be small at moderate densities $na^2\sim 0.1$. However, the condition \textbf{(ii)} is clearly violated, which means that a proper description of scattering would require going beyond the two-channel model adopted here. We expect that the phenomenology of resonant scattering would still persist at $\hbar\Omega_L\sim \mu_B g_L B$, albeit in a strongly modified form. We postpone the corresponding analysis for future work. We also expect, that in near future either TMD- or perovskite-based settings would provide favourable conditions.   

\section{Conclusions}

In summary, we have explored a possibility of controllable interactions of identical electromagnetic bosons (excitons or polaritons) in strained 2D semiconductors. The strain induces an effective in-plane magnetic field which couples the spin-polarized two-body scattering continua to a singlet biexciton state, thus acting as a coherent link transforming the biexciton into a scattering resonance. Application of an ordinary transverse magnetic field yields the phenomenology of the Feshbach resonance, wherein the lower-energy "open" scattering channel is tuned across the resonant biexciton level ("closed" channel). For a broad resonance, the two-body interaction of bosons changes from increasingly strong repulsion to attraction, as one crosses the resonance, in full analogy with the control of interactions achieved in ultra-cold atoms. \cite{Chin2010} In contrast to the earlier proposals,\cite{Wouters2007, carusotto2010, Andreev2016, Andreev2020, Kuhlenkamp2022} here one has direct access to the width of the resonance through the strain engineering of the host heterostructure. The resonance width defines, in particular, the timescale $\mathcal{T}_\alpha$ [Eq. \eqref{SqueezeTime} with $\alpha=X$ or $\alpha=L$ for excitons or lower polaritons, respectively] characterizing efficiency of squeezing of the electromagnetic bosons at unitarity.\cite{Andreev2020} We expect the effect to be significant in the currently available samples of TMD's.

From the non-linear optics perspective, the scattering resonance enables simultaneously tuneable third-order [$\chi^{3}$] and an analog second-order [$\chi^{2}$] susceptibilities. Interestingly, the latter does not require broken inversion symmetry of the medium. The ensuing parametric processes for macroscopically populated bosonic modes may be conveniently regarded as Bose-stimulated reactions of fusion and disintegration. These reactions occur on the characteristic timescale $\mathcal{T}_\alpha$ and are accompanied by squeezing of the emitted light.   

For excitons, we find that the scattering amplitude [Eq. \eqref{ResAmplitude}] possesses a double-pole structure: in sufficiently strong transverse magnetic field a resonance coexists with a synthetic bound state.  The synthetic molecule consists of a singlet biexciton core and a spin-polarized quantum halo which extends far beyond the range of the microscopic interaction potential between the excitons. The halo has giant oscillator strength \cite{Rashba1962} which is expected to grow exponentially with the applied magnetic field. The spin state of the molecule can be adiabatically tuned from a separable state (halo) to a maximally entangled Bell state (biexciton) by sweeping the magnetic field across the resonance. Our study thus hints explicitly at a connection between the entanglement and squeezing of resonantly paired photons, also anticipated from a general standpoint \cite{Scully}, and which yet remains to be explored.

\begin{acknowledgements} 
I acknowledge V. Shatokhin and A. Buchleitner for clarifying remarks. 
\end{acknowledgements}

\appendix
\section{Background polariton interactions}
\label{AnnexeA}

In the ultra-cold limit, where the de Broglie wavelength of the particle relative motion $2\pi/k$ greatly exceeds the microscopic range $R_{\sigma\sigma^\prime}$ of the interaction potential $V_{\sigma\sigma^\prime}^{(X)}(r)$, the scattering becomes fully characterized by the 2D scattering length $a_{\sigma\sigma^\prime}^{(X)}$. Simple approximations for $V_{\sigma\sigma^\prime}^{(X)}(r)$ may be used to establish the qualitative dependence of $a_{\sigma\sigma^\prime}^{(X)}$ on $R_{\sigma\sigma^\prime}$ and on the average magnitude of $V_{\sigma\sigma^\prime}^{(X)}(r)$. We shall assume that the Fourier transform $V_{\sigma\sigma^\prime}^{(X)}(\bm{k}^\prime-\bm{k})=(2\pi)^{-2}\int V_{\sigma\sigma^\prime}^{(X)}(r)e^{i(\bm{k}^\prime-\bm{k})\bm{r}}d\bm{r}$ exists and replace it by the separable force   
\begin{equation}
\label{Force}
V_{\sigma\sigma^\prime}^{(X)}(\bm{k}^\prime,\bm{k})=\frac{\hbar^2}{m_X} \lambda_{\sigma\sigma^\prime}e^{-(k^\prime R_{\sigma\sigma^\prime})^2/2}e^{-(kR_{\sigma\sigma^\prime})^2/2}.
\end{equation}
By using the Lippmann-Schwinger equation for the two-body $T$-matrix
\begin{equation}
\label{Lippmann}
T_{\sigma\sigma^\prime}^{(\alpha)}(E_{\bm{k}}^{(\alpha)}+i0)=V_{\sigma\sigma^\prime}^{(\alpha)}+V_{\sigma\sigma^\prime}^{(\alpha)}G_0^{(\alpha)}(E_{\bm{k}}^{(\alpha)}+i0) T_{\sigma\sigma^\prime}^{(\alpha)}
\end{equation}
with $G_0^{\alpha}(E_{\bm{k}}^{(\alpha)}+i0)$ being the Green function for the free relative motion at the energy $E_{\bm{k}}^{(\alpha)}\equiv\hbar^2k^2/m_\alpha$ and putting $\alpha=X$ (excitons), one immediately gets Eq. \eqref{Amplitudes} with
\begin{equation}
\label{ExcitonScattLength}
a_{\sigma\sigma^\prime}^{(X)}=R_{\sigma\sigma^\prime}e^{[\gamma/2-1/(2\pi \lambda_{\sigma\sigma^\prime})]}.
\end{equation}
 The magnitudes $\lambda_{\uparrow\uparrow}$ and $\lambda_{\downarrow\downarrow}$ of the repulsive potentials are not fixed. In the limit $\lambda_{\sigma\sigma}\rightarrow +\infty$ one recovers the well-known result for the scattering off an impenetrable disk of radius $2R_{\sigma\sigma}e^{-\gamma/2}$. In the opposite limit $\lambda_{\sigma\sigma}\rightarrow 0$ one gets the Born approximation $f_{\sigma\sigma}^{(X)}(k)=2\pi^2\lambda_{\sigma\sigma}$. The strength of the attractive force $\lambda_{\uparrow\downarrow}<0$ and its range $R_{\uparrow\downarrow}$  should be chosen such as to get the bound state with energy $\varepsilon$ and radius $\sim\hbar/\sqrt{m_X \varepsilon}$. By putting $R_{\uparrow\downarrow}\equiv\hbar/\sqrt{m_X \varepsilon}$ and substituting Eq. \eqref{Force} into the Schrodinger Eq. \eqref{Biexciton} written in the $\bm{k}$-space, we obtain
\begin{equation*}
\phi(\bm q)(\varepsilon-\varepsilon_{\bm q})=\lambda_{\uparrow\downarrow}e^{-\varepsilon_{\bm q}/2|\varepsilon|}\frac{(2\pi)^2}{S}\sum\limits_{\bm k}\phi(\bm k)e^{-\varepsilon_{\bm k}/2|\varepsilon|},
\end{equation*}
where $\phi(\bm k)$ is the Fourier transform of Eq. \eqref{Gaussian}. This further yields
\begin{equation*}
1=\frac{(2\pi)^2\lambda_{\uparrow\downarrow}}{S}\sum\limits_{\bm k}\frac{e^{-\varepsilon_{\bm k}/|\varepsilon|}}{\varepsilon-\varepsilon_{\bm k}},
\end{equation*}
so that one obtains
\begin{equation}
\lambda_{\uparrow\downarrow}=[\pi e \mathrm{Ei}(-1)]^{-1}
\end{equation}
with $\mathrm{Ei}(x)$ being the exponential integral function.

The polariton interaction potential reads $V_{\sigma\sigma^\prime}^{(L)}(\bm{k}^\prime,\bm{k})=X^2_{\bm{k}^\prime}X^2_{\bm{k}}V_{\sigma\sigma^\prime}^{(X)}(\bm{k}^\prime,\bm{k})$. Consistently, we look for the polariton $T$-matrix in the form
\begin{equation}
\label{PolaritonTmatrix}
\begin{split}
&T_{\sigma\sigma^\prime}^{(L)}(\bm{k}^\prime,\bm{k},E_{\bm{k}}^{(L)}+i0)=\\
&\frac{\hbar^2}{m_X}t_{\sigma\sigma^\prime}^{(L)}(k)X^2_{\bm{k}^\prime}X^2_{\bm{k}}e^{-(k^\prime R_{\sigma\sigma^\prime})^2/2}e^{-(kR_{\sigma\sigma^\prime})^2/2}.
\end{split}
\end{equation}           
By substituting this \textit{ansatz} into Eq. \eqref{Lippmann} with $\alpha=L$ (lower polaritons), we obtain
\begin{equation}
\label{PolaritonSmallTmatrix}
t_{\sigma\sigma^\prime}^{(L)}(k)=[\lambda_{\sigma\sigma^\prime}^{-1}-\Pi_{\sigma\sigma^\prime}^{(L)}(E_{\bm{k}}^{(L)}+i0)]^{-1},
\end{equation}
where the polarization bubble
\begin{widetext}
\begin{equation}
\Pi_{\sigma\sigma^\prime}^{(L)}(E_{\bm{k}}^{(L)}+i0)=\pi\int\limits_{0}^{E_0}\frac{X_0^4 e^{-E/\varepsilon_{\sigma\sigma^\prime}}}{E_{\bm{k}}^{(L)}-m_X/m_L E+i0}dE+\pi\int\limits_{E_0}^{+\infty}\frac{X_{+\infty}^4 e^{-E/\varepsilon_{\sigma\sigma^\prime}}}{E_{\bm{k}}^{(L)}-(\hbar\Omega_R+E)+i0}dE
\end{equation}
\end{widetext}
has been split into two parts corresponding to the photon-like and exciton-like regions of the polariton dispersion. Here we have defined $E_0\equiv  \hbar\Omega_R m_L/m_X$ and $\varepsilon_{\sigma\sigma^\prime}\equiv \hbar^2/m R_{\sigma\sigma^\prime}^2$. By evaluating the integrals we find
\begin{widetext}
\begin{equation}
\label{PolarizationBubble}
\Pi_{\sigma\sigma^\prime}^{(L)}(E_{\bm{k}}^{(L)}+i0)=\pi X_0^4 \frac{m_L}{m_X}\left(\ln\left[\frac{E_{\bm{k}}^{(L)}}{\hbar\Omega_R-E_{\bm{k}}^{(L)}}\right]-i\pi\right)+\pi X_{+\infty}^4e^{(\hbar\Omega_R-E_{\bm{k}}^{(L)})/\varepsilon_{\sigma\sigma^\prime}}\mathrm{Ei}\left(\frac{E_{\bm{k}}^{(L)}-\hbar\Omega_R}{\varepsilon_{\sigma\sigma^\prime}}\right)
\end{equation}
\end{widetext}
for $E_{\bm{k}}^{(L)}<\hbar\Omega_R$ and
\begin{widetext}
\begin{equation}
\Pi_{\sigma\sigma^\prime}^{(L)}(E_{\bm{k}}^{(L)}+i0)=\pi X_0^4 \frac{m_L}{m_X}\ln\left[\frac{E_{\bm{k}}^{(L)}}{\hbar\Omega_R-E_{\bm{k}}^{(L)}}\right]+\pi X_{+\infty}^4e^{(\hbar\Omega_R-E_{\bm{k}}^{(L)})/\varepsilon_{\sigma\sigma^\prime}}\left[\mathrm{Ei}\left(\frac{E_{\bm{k}}^{(L)}-\hbar\Omega_R}{\varepsilon_{\sigma\sigma^\prime}}\right)-i\pi\right]
\end{equation}
\end{widetext}
for $E_{\bm{k}}^{(L)}>\hbar\Omega_R$. The low-energy \textit{polariton} scattering corresponds to $E_{\bm{k}}^{(L)}<\hbar\Omega_R$.  For repulsive potentials with $\lambda_{\sigma\sigma}>0$ one obtains by letting $E_{\bm{k}}^{(L)}\rightarrow 0$ in  Eq. \eqref{PolaritonTmatrix} and assuming $\varepsilon_{\sigma\sigma^\prime}\gg \hbar\Omega_R$ 
\begin{equation}
\label{RepulsivePolaritonTmatrix}
T_{\sigma\sigma}^{(L)}=-\frac{\hbar^2}{m_X}\frac{(2\pi)^{-1}X_0^4}{\ln{\left[k_R a_{\sigma\sigma}^{(L)}\right]}}, 
\end{equation}
where $k_R=\sqrt{m_L \Omega_R/\hbar}$ and  $a_{\sigma\sigma}^{(L)}=\sqrt{m_X/m_L} a_{\sigma\sigma}^{(X)}$ with $a_{\sigma\sigma}^{(X)}$ given by Eq. \eqref{ExcitonScattLength}. This result is in agreement with that obtained in the earlier work.\cite{Bleu2020} For the attractive potential $\lambda_{\uparrow\downarrow}<0$ the situation is more subtle. Here $\varepsilon_{\uparrow\downarrow}\equiv |\varepsilon| \sim \hbar\Omega_R$ and one cannot take advantage of the logarithmic expansion of $\mathrm{Ei}(x)$ at $x\rightarrow 0$ in the second term of Eq. \eqref{PolarizationBubble} to obtain the formula similar to Eq. \eqref{RepulsivePolaritonTmatrix}. Instead, as $\hbar\Omega_R\rightarrow|\varepsilon|$ and $E_{\bm{k}}^{(L)}\rightarrow 0$, this term cancels with the $\lambda_{\uparrow\downarrow}^{-1}$ term in Eq. \eqref{PolaritonSmallTmatrix} and we are left with
\begin{equation}
\label{AttractivePolaritonTmatrix}
T_{\uparrow\downarrow}^{(L)}(k)=-\frac{\hbar^2}{m_L}\frac{(2\pi)^{-1}}{\ln[ka_{\uparrow\downarrow}^{(L)}]-i\pi/2}, 
\end{equation}
where 
\begin{equation}
a_{\uparrow\downarrow}^{(L)}=\sqrt{m_X/m_L}\exp\left[\frac{m_X}{m_L}\frac{\hbar\Omega_R-|\varepsilon|}{2\beta_{\uparrow\downarrow}}\right] a,
\end{equation}
$\beta_{\uparrow\downarrow}=X_0^4|\varepsilon|/(1+\lambda_{\uparrow\downarrow})\sim |\varepsilon|$ and $a\equiv\hbar/\sqrt{m_X |\varepsilon |}$. The polariton scattering amplitude reads
\begin{equation}
\bar f_{\uparrow\downarrow}^{(L)}(k)=\frac{\pi}{\ln[ka_{\uparrow\downarrow}^{(L)}]-i\pi/2}.
\end{equation}
Since typically $m_L\ll m_X$, for $\hbar\Omega_R>|\varepsilon|$, \textit{i.e.} when the discrete level is embedded into the continuum of the polariton scattering states, we obtain the scattering off a weakly-bound state. In 3D such scattering would produce increasingly strong repulsive interaction. In 2D, in contrast, the real part of the scattering amplitude vanishes as $k^{-1}$ approaches $a_{\uparrow\downarrow}^{(L)}$. For $\hbar\Omega_R\ll |\varepsilon|$ the expression for $T_{\uparrow\downarrow}^{(L)}$  takes the from of Eq. \eqref{RepulsivePolaritonTmatrix}. In the intermediate region $0<|\varepsilon|-\hbar\Omega_R\ll |\varepsilon|$ the effective interaction  $g_{\uparrow\downarrow}^{(L)}\equiv (2\pi)^2 T_{\uparrow\downarrow}^{(L)}$ is positive and somewhat larger than the corresponding interaction constants in the nominally repulsive channels $g_{\sigma\sigma}^{(L)}\equiv (2\pi)^2 T_{\sigma\sigma}^{(L)}$. Within our simple model we do not find the phenomenology of the resonant scattering in the bare interaction of polaritons with opposite spins.

\section{Resonant scattering amplitude}
\label{AnnexeB}

The resonant part of the boson scattering amplitude reads
\begin{equation}
\label{fres}
f_{\uparrow\uparrow, \mathrm{res}}^{(\alpha)}(k)=-\frac{2m_\alpha|\Omega_{\uparrow\downarrow}^{(\alpha)}|^2}{\hbar^2}\frac{\pi^2 |\braket{\varphi|\bar\psi_{\uparrow\uparrow,\bm{k}}^{(\alpha)}}|^2}{\hbar^2 k^2/m_\alpha-\bar\Delta_\alpha-\Pi_\alpha(E^{(\alpha)}_{\bm{k}}+i0)},
\end{equation}
where $\bar\Delta_\alpha$ is the bare detuning governed by the external magnetic field. The key elements of Eq. \eqref{fres} are the polarization bubble
\begin{equation}
 \Pi_\alpha(E^{(\alpha)}_{\bm{k}}+i0)=\int \frac{2 |\Omega_{\uparrow\downarrow}^{(\alpha)}|^2 |\braket{\bm{q}|\varphi}|^2}{E^{(\alpha)}_{\bm{k}}-E^{(\alpha)}_{\bm{q}}+i0}d\bm{q}
 \end{equation}
 and the overlap of the background stationary scattering state $\bar\psi_{\uparrow\uparrow,\bm{k}}^{(\alpha)}(\bm{r})$ with the bare molecule $\varphi(r)$.   

Let us evaluate Eq. \eqref{fres} for the lower polaritons ($\alpha=L$). As in the case of background polariton interactions (see above), the integration in the bubble $\Pi_L(E^{(L)}_{\bm{k}}+i0)$ may be split into two parts corresponding to the photon-like and exciton-like regions of the polariton dispersion:
\begin{widetext}
\begin{equation}
\Pi_L(y+i0)=\frac{2|\Omega_{\uparrow\downarrow}^{(L)}|^2}{E_a^{(L)}}\left (\int\limits_0^{\frac{\hbar\Omega_R}{E_a^{(L)}}}\frac{e^{-x}}{y-x+i0}dx+\int\limits_\frac{\hbar\Omega_R}{E_a^{(L)}}^{\infty}\frac{e^{-x}}{y-\frac{m_L}{m_X}x-\frac{\hbar\Omega_R}{E_a^{(L)}}}dx\right),
\end{equation}
\end{widetext}
where $E_a^{(L)}\equiv\hbar^2/m_L a^2$ and $y\equiv E^{(L)}_{\bm{k}}/E_a^{(L)}$. By our assumption, $\hbar\Omega_R\sim|\varepsilon|\sim\hbar^2/m_X a^2$, and we can write $\hbar\Omega_R/E_a^{(L)}\approx m_L/m_X\equiv \sigma\ll 1$. We then notice that $y\ll\sigma$ and obtain
\begin{widetext}
\begin{equation}
\begin{split}
\Pi_L(y+i0)&=\frac{2|\Omega_{\uparrow\downarrow}^{(L)}|^2}{E_a^{(L)}}\left [e^{-y}[\mathrm{Ei}(y)-\mathrm{Ei}(-\sigma+y)-i\pi]+\frac{e^{1-y/\sigma}}{\sigma}\mathrm{Ei}(-1-\sigma+y/\sigma)\right]\\
&=\frac{2|\Omega_{\uparrow\downarrow}^{(L)}|^2}{E_a^{(X)}}e\mathrm{Ei}(-1)+\frac{2|\Omega_{\uparrow\downarrow}^{(L)}|^2}{E_a^{(L)}}\ln\left(\frac{E^{(L)}_{\bm{k}}}{E_a^{(X)}}\right).
\end{split}
\end{equation}
\end{widetext}
By substituting this result into Eq. \eqref{fres} we find
\begin{equation}
f_{\uparrow\uparrow, \mathrm{res}}^{(L)}(k)=\frac{\pi \wp_L}{(-\hbar^2k^2/m_L+\Delta_L)/2\beta_L+\ln(ka_L)-i\pi/2},
\end{equation}
where $\beta_L\equiv 2m_L a^2 |\Omega_{\uparrow\downarrow}^{(L)}|^2/\hbar^2$, $\Delta_L\equiv\bar\Delta_L+\sigma^{-1}\beta_L e\mathrm{Ei}(-1)$, $a_L\equiv i a/\sqrt{\sigma}$ and $\wp_L\equiv \pi/a^2 |\braket{\varphi|\bar\psi_{\uparrow\uparrow,\bm{k}}^{(L)}}|^2$. According to Eq. \eqref{PolaritonTmatrix}, the background polariton scattering amplitude $\bar f_{\uparrow\uparrow}^{(L)}(k)\equiv-(2\pi)^2 m_L/2\hbar^2 T^{(L)}_{\uparrow\uparrow}(k) $ is proportional to $\sigma=m_L/m_X\ll 1$. Hence, we may let $\bar\psi_{\uparrow\uparrow,\bm{k}}^{(L)}(\bm{r})=e^{i\bm{k}\cdot\bm{r}}/2\pi$ and obtain $\wp_L=1$.

\section{Distorsion of the exciton continuum due to the background scattering}
\label{AnnexeC}

Let us now evaluate the parameter
\begin{equation}
\wp_X\equiv \pi/a^2|\braket{\varphi|\bar\psi_{\uparrow\uparrow,\bm{k}}^{(X)}}|^2
\end{equation}
which accounts for the distortion of the exciton continuum due to the background potential $V_{\uparrow\uparrow}^{(X)}(r)$. To this end, we decompose the stationary scattering state $\bar\psi_{\uparrow\uparrow,\bm{k}}^{(X)}(\bm r)$ into the partial waves,
\begin{equation*}
\bar\psi_{\uparrow\uparrow,\bm{k}}^{(X)}(\bm r)=\frac{1}{2\pi}\sum\limits_{m=-\infty}^{+\infty}i^m \bar\psi_{\uparrow\uparrow, m, k}^{(X)}(r)e^{im\varphi},
\end{equation*}
and take the impenetrable disk \textit{ansatz} for the $s$-wave ($m=0$):
\begin{equation*}
\begin{split}
&\bar\psi_{\uparrow\uparrow, 0, k}^{(X)}(r)\equiv 0,\:\:\:\:r\leqslant R_{\uparrow\uparrow}^{(X)}\\
&\bar\psi_{\uparrow\uparrow, 0, k}^{(X)}(r)=\\
&\frac{1}{2}[H_0^{-}(kr)+(1+i\bar f_{\uparrow\uparrow}^{(X)}(k))H_0^{+}(kr)],\:\:\:r> R_{\uparrow\uparrow}^{(X)},
\end{split}
\end{equation*}
where $H_0^{\pm}(kr)$ are the Hankel functions and the background scattering amplitude $\bar f_{\uparrow\uparrow}^{(X)}(k)$ is given by Eq. \eqref{Amplitudes} with $a_{\uparrow\uparrow}^{(X)}\equiv R_{\uparrow\uparrow}^{(X)}$. By using the Gaussian biexciton wave function [Eq. \eqref{Gaussian}] and assuming $R_{\uparrow\uparrow}^{(X)}\ll a$, we obtain in the limit $k\rightarrow 0$:
\begin{equation*}
\braket{\varphi|\bar\psi_{\uparrow\uparrow,\bm{k}}^{(X)}}\approx\frac{a}{\sqrt{\pi}}\frac{\ln\left(\tfrac{R_{\uparrow\uparrow}^{(X)}e^{\gamma/2}}{2a}\right)}{\ln\left(\tfrac{kR_{\uparrow\uparrow}^{(X)}e^{\gamma}}{2}\right)},
\end{equation*}
which upon sending $R_{\uparrow\uparrow}^{(X)}/a\rightarrow 0$ yields  Eq. \eqref{Factor} for $\wp_X$.
  
\section{Entanglement degree of the molecular polarization}
\label{AnnexeD}

By tracing out the spatial parts of the wavefunction \eqref{Halo}, we obtain the polarization density matrix
\begin{equation}
\label{DensityMatrix}
\hat \rho=\left( \begin{array}{cccc}
\braket{\Upsilon|\Upsilon} & \frac{w}{\sqrt{2}}\braket{\varphi|\Upsilon} &  \frac{w}{\sqrt{2}}\braket{\varphi|\Upsilon} & 0 \\
\frac{w}{\sqrt{2}}\braket{\Upsilon|\varphi} & \frac{w^2}{2}\braket{\varphi|\varphi} & \frac{w^2}{2}\braket{\varphi|\varphi} & 0\\
\frac{w}{\sqrt{2}}\braket{\Upsilon|\varphi} & \frac{w^2}{2}\braket{\varphi|\varphi} &\frac{w^2}{2}\braket{\varphi|\varphi} & 0\\
0 & 0 & 0 & 0 
\end{array} \right),
\end{equation}
written in the "computational" basis $\ket{\uparrow\uparrow}$, $\ket{\uparrow\downarrow}$, $\ket{\downarrow\uparrow}$ and $\ket{\downarrow\downarrow}$. We have used the notations $\Upsilon(r)=\braket{\bm{r}|\Upsilon}$ and $\varphi(r)=\braket{\bm{r}|\varphi}$. One has 
\begin{equation}
\braket{\Upsilon|\Upsilon}=1-w^2
\end{equation}
and $\braket{\varphi|\varphi}=1$, so that the mixed state \eqref{DensityMatrix} is fully characterized by two parameters: the relative weight of the core $w^2$ and the overlap integral $\braket{\varphi|\Upsilon}$. Surprisingly, as we show below, the latter drops from the final result for the entanglement measure.  

Let us evaluate Eq. \eqref{concurrence} by following the analytical prescription \cite{Wootters1998, Mintert2005} based on the methods of linear algebra. To this end, we diagonalize $\hat\rho$ and build an auxiliary symmetric matrix $\hat\tau$ with the elements
\begin{equation}    
\tau_{+-}\equiv \sqrt{\lambda_+\lambda_-}\braket{-|\hat\sigma_y\otimes\hat\sigma_y|+},
\end{equation}
where
\begin{equation}
\label{eigenvalues}
\lambda_{\pm}=\frac{1}{2}\left [1\pm\sqrt{(1-2w^2)^2+4w^2\braket{\varphi|\Upsilon}}\right]
\end{equation}
are the non-zero eigenvalues of $\hat\rho$ and
\begin{equation}
\ket{\pm}=\frac{1}{\sqrt{2+\nu_\pm^2}}\left (\begin{array}{c}
\nu_\pm\\
1\\
1\\
0
\end{array}\right)
\end{equation}
are the corresponding (normalized) eigenvectors, with
\begin{equation}
\label{eigenvectors}
\nu_\pm\equiv \frac{1-2w^2\pm\sqrt{(1-2w^2)^2+4w^2\braket{\varphi|\Upsilon}}}{\sqrt{2}w\braket{\varphi|\Upsilon}}.
\end{equation}
We then compute the singular values of $\hat\tau$ as square roots of the eigenvalues of the positive hermitian matrix $\hat\tau\hat\tau^\dagger$. The unique non-zero singular value reads
\begin{equation}
\mathcal L=2\left (\frac{\lambda_+}{2+\nu_+^2}+\frac{\lambda_-}{2+\nu_-^2}\right),
\end{equation}
which, upon substitution of Eq. \eqref{eigenvalues} and Eq. \eqref{eigenvectors}, yields the formula
\begin{equation}
c[\hat\rho]\equiv \mathcal L=w^2
\end{equation}  
for the concurrence of the mixed state $\hat\rho$. The absence of the overlap integral $\braket{\varphi|\Upsilon}$ in this final result is surprising and calls for further investigation.

\section{Boltzmannian gas of molecules}
\label{AnnexeE}

At low molecular densities ($n a_\gamma^2\ll 1$) the polarization density matrix $\hat\rho$ defined by Eq. \eqref{PolDensMatrix} may be reconstructed by performing cross correlation polarization measurements in a photon coincidence setup analogous to that employed previously for QD's. \cite{Stevenson2006} Indeed, assuming the center-of-mass (c.o.m.) motion of the molecules being uncorrelated with their internal state [represented by the pure state $\Psi(\bm r)$], the total density matrix for a grand-canonical ensemble may be written in a factorized form
\begin{equation}
\label{GrandCan}
\hat\rho_\mathrm{GC}=\frac{e^{\mu \hat N/k_B T}}{Z_\mathrm{GC}} (e^{-\hat {\bm K}^2\lambda_T^2})^{\otimes N}\otimes\hat\rho^{\otimes N}
\end{equation}
with
\begin{equation}
Z_\mathrm{GC}=\sum\limits_N \frac{e^{\mu \hat N/k_B T}}{N!}(\mathrm{Tr}[e^{-\hat {\bm K}^2\lambda_T^2}])^N
\end{equation}
being the grand canonical partition function and the thermal de Broglie wavelength of a molecule $\lambda_T\equiv\sqrt{\hbar^2/4mk_BT}$ satisfying the condition 
\begin{equation}
a_\gamma\ll \lambda_T \ll n^{-1/2}.
\end{equation}
A two-photon coincidence count randomly picks out a pair from the total flux and projects its polarization state onto a specific measurement state $\ket{\psi_\nu}$ set by the corresponding optical elements (polarizers, quarter- and half-wave plates). The average number of such coincidences $n_\nu$ should include projection of the grand canonical ensemble \eqref{GrandCan} onto a single pair polarization state $\hat\rho$ and consecutive averaging over the total number of molecules $N$, as well as over their c.o.m. positions $\bm R_i$. Formally, such a procedure may be written as
\begin{equation}
\label{Counts}
n_\nu=\mathcal{N}\bra{\psi_\nu}\sum\limits_N\braket{\bm R_1,...\bm R_N|\hat {\mathcal P}^\dagger \hat\rho_\mathrm{GC} \hat {\mathcal P} | \bm R_1,...\bm R_N}\ket{\psi_\nu},
\end{equation}
where
\begin{equation}
\hat {\mathcal P}\equiv\frac{1}{N}\sum\limits_{i=1}^{N}\sum_{\sigma,\sigma^\prime}\ket{\sigma\sigma^\prime}_i\bra{\sigma^\prime\sigma}
\end{equation}
is the required projector, $\mathcal{N}$ is a constant which depends on the photoluminescence yield and detector efficiency, and $\nu=1,..16$ labels a particular measurement state within (what would be) a tomographically complete set of measurements.\cite{James2001} By substituting Eq. \eqref{GrandCan} into Eq. \eqref{Counts}, one obtains
\begin{equation}
n_\nu=\mathcal{N}\bra{\psi_\nu}\hat\rho \ket{\psi_\nu}.
\end{equation}
Hence, the single-molecule density matrix $\hat\rho$ does indeed constitute a legitimate observable under the conditions formulated above.
 
For a quantum degenerate gas, the description in terms of a pure state \eqref{Halo} is no longer adequate and one should proceed from a many-body solution of the Hamiltonian \eqref{SecondHamiltonian}.    
                 
\bibliography{References}
\end{document}